\documentclass[a4paper,11pt]{article}
\pdfoutput=1
\usepackage{./auxi/jheppub}
\usepackage[utf8]{inputenc}

\usepackage{slashed} 
\usepackage{mathtools} 
\usepackage{tikz}
\usepackage{dsfont} 
\usetikzlibrary{positioning,arrows,calc}
\usetikzlibrary{arrows.meta}
\usetikzlibrary{decorations.pathmorphing}
\usetikzlibrary{decorations.markings}
\usetikzlibrary{shapes.geometric}
\usetikzlibrary{patterns}
\usepackage{xparse}
\usepackage{diagbox} 
\usepackage{tensor} 
\usepackage{subcaption} 

\DeclareMathOperator{\Disc}{Disc}

\newcommand{\vvev}[1]{\langle\!\langle\, #1 \, \rangle\!\rangle}
\newcommand{\vev}[1]{\langle\, #1 \, \rangle}

\newcommand{\Op}{\mathcal{O}}

\newcommand{\uh}{\hat{u}}

\newcommand{\Dh}{\hat{\Delta}}

\newcommand{\Wl}{\mathcal{W}_\ell}

\newcommand{\hh}{\hat{h}}
\newcommand{\fh}{\hat{f}}

\newcommand{\Gh}{\hat{\mathcal{G}}}

\newcommand{\Oh}{\hat{\mathcal{O}}}

\newcommand{\OR}{\sigma}

\def\Am{{\mathcal{A}}}
\def\Bm{{\mathcal{B}}}
\def\Cm{{\mathcal{C}}}

\def\Fm{{\mathcal{F}}}
\def\Gm{{\mathcal{G}}}

\def\Nm{{\mathcal{N}}}
\def\Om{{\mathcal{O}}}
\def\Pm{{\mathcal{P}}}

\def\Wm{{\mathcal{W}}}

\usepackage{bbm}

\newcommand\yb{{\bar{y}}}
\newcommand\zb{{\bar{z}}}

\newif\ifstartcompletesineup
\newif\ifendcompletesineup
\pgfkeys{
    /pgf/decoration/.cd,
    start up/.is if=startcompletesineup,
    start up=true,
    start up/.default=true,
    start down/.style={/pgf/decoration/start up=false},
    end up/.is if=endcompletesineup,
    end up=true,
    end up/.default=true,
    end down/.style={/pgf/decoration/end up=false}
}
\pgfdeclaredecoration{complete sines}{initial}
{
    \state{initial}[
        width=+0pt,
        next state=upsine,
        persistent precomputation={
            \ifstartcompletesineup
                \pgfkeys{/pgf/decoration automaton/next state=upsine}
                \ifendcompletesineup
                    \pgfmathsetmacro\matchinglength{
                        0.5*\pgfdecoratedinputsegmentlength / (ceil(0.5* \pgfdecoratedinputsegmentlength / \pgfdecorationsegmentlength) )
                    }
                \else
                    \pgfmathsetmacro\matchinglength{
                        0.5 * \pgfdecoratedinputsegmentlength / (ceil(0.5 * \pgfdecoratedinputsegmentlength / \pgfdecorationsegmentlength ) - 0.499)
                    }
                \fi
            \else
                \pgfkeys{/pgf/decoration automaton/next state=downsine}
                \ifendcompletesineup
                    \pgfmathsetmacro\matchinglength{
                        0.5* \pgfdecoratedinputsegmentlength / (ceil(0.5 * \pgfdecoratedinputsegmentlength / \pgfdecorationsegmentlength ) - 0.4999)
                    }
                \else
                    \pgfmathsetmacro\matchinglength{
                        0.5 * \pgfdecoratedinputsegmentlength / (ceil(0.5 * \pgfdecoratedinputsegmentlength / \pgfdecorationsegmentlength ) )
                    }
                \fi
            \fi
            \setlength{\pgfdecorationsegmentlength}{\matchinglength pt}
        }] {}
    \state{downsine}[width=\pgfdecorationsegmentlength,next state=upsine]{
        \pgfpathsine{\pgfpoint{0.5\pgfdecorationsegmentlength}{0.5\pgfdecorationsegmentamplitude}}
        \pgfpathcosine{\pgfpoint{0.5\pgfdecorationsegmentlength}{-0.5\pgfdecorationsegmentamplitude}}
    }
    \state{upsine}[width=\pgfdecorationsegmentlength,next state=downsine]{
        \pgfpathsine{\pgfpoint{0.5\pgfdecorationsegmentlength}{-0.5\pgfdecorationsegmentamplitude}}
        \pgfpathcosine{\pgfpoint{0.5\pgfdecorationsegmentlength}{0.5\pgfdecorationsegmentamplitude}}
}
    \state{final}{}
}

\definecolor{bgbox}{RGB}{255,254,230}
\definecolor{setupplane}{RGB}{230,230,230}
\definecolor{gluoncolor}{RGB}{207,54,108}
\definecolor{vertexcolor}{RGB}{53,152,219}
\definecolor{SEcolor}{RGB}{176,156,255}
\definecolor{blobcolor}{RGB}{190,180,230}
\tikzset{
corner/.style={line width=1pt,dashed,draw=black,dash pattern=on 6pt off 4pt},
scalar/.style={line width=1pt,draw=black},
gluon/.style={line width=1pt,decorate, draw=gluoncolor,
    decoration={complete sines,aspect=0,amplitude=1.25mm,segment length=1.5mm,start up,end up}},
ghost/.style={line width=1pt,loosely dotted,draw=black},
wilson/.style={line width=2pt,draw=black},
 }
\NewDocumentCommand\semiloop{O{black}mmmO{}O{above}}
{%
\draw[#1] let \p1 = ($(#3)-(#2)$) in (#3) arc (#4:({#4+180}):({0.5*veclen(\x1,\y1)})node[midway, #6] {#5};)
}

\usepackage{simplewick}
\usepackage{dsfont}
\usepackage{float}
\usepackage{pgfplots}
\pgfplotsset{compat=1.14}

\let\oldbfseries=\bfseries
\let\oldmdseries=\mdseries
\let\oldnormalfont=\normalfont
\renewcommand{\bfseries}{\oldbfseries\boldmath}
\renewcommand{\mdseries}{\oldmdseries\unboldmath}
\renewcommand{\normalfont}{\oldnormalfont\unboldmath}

\makeatletter
\newlength{\apb@width}
\newcommand{\autoparbox}[2][c]{\settowidth{\apb@width}{#2}\parbox[#1]{\apb@width}{#2}}

\makeatother

\DeclareMathOperator{\tr}{tr}

\def\Am{{\mathcal{A}}}
\def\Bm{{\mathcal{B}}}
\def\Cm{{\mathcal{C}}}

\def\Fm{{\mathcal{F}}}
\def\Gm{{\mathcal{G}}}

\def\Nm{{\mathcal{N}}}
\def\Om{{\mathcal{O}}}
\def\Pm{{\mathcal{P}}}

\def\zb{{\bar{z}}}

\def\r{\rho}

\newcommand{\beq}{\begin{equation}}
\newcommand{\eeq}{\end{equation}}

\definecolor{nicegreen}{rgb}{0.1,0.6,0.1}

\newmuskip\pFqskip
\mathchardef\pFcomma=\mathcode`,

\makeatletter
\renewcommand*\env@matrix[1][\arraystretch]{%
  \edef\arraystretch{#1}%
  \hskip -\arraycolsep
  \let\@ifnextchar\new@ifnextchar
  \array{*\c@MaxMatrixCols c}}
\makeatother

\allowdisplaybreaks 

\title{\center{Bootstrapping holographic defect correlators in $\mathcal{N}=4$ super Yang-Mills}}

\author[1]{Julien Barrat,}
\author[2]{Aleix Gimenez-Grau,}
\author[2]{Pedro Liendo.}

\affiliation[1]{Institut für Physik und IRIS Adlershof, Humboldt-Universität zu Berlin,
 Zum Großen Windkanal 2, 12489 Berlin, Germany}
\affiliation[2]{DESY Hamburg, Theory Group, Notkestra{\ss}e 85, D-22607 Hamburg, Germany}

\emailAdd{julien.barrat@hu-berlin.de, aleix.gimenez@desy.de, pedro.liendo@desy.de}

\preprint{HU-EP-21/29-RTG, DESY 21-133}

\abstract{We study two-point functions of single-trace half-BPS operators in the presence of a supersymmetric Wilson line in $\mathcal{N}=4$ SYM. We use inversion formula technology in order to reconstruct the CFT data starting from a single discontinuity of the correlator. In the planar strong coupling limit only a finite number of conformal blocks contributes to the discontinuity, which allows us to obtain elegant closed-form expressions for two-point functions of single-trace operators $\Om_J$ of weight $J=2,3,4$.  Our final result passes a number of non-trivial consistency checks: it has the correct discontinuity, it satisfies the superconformal Ward identities, it has a sensible expansion in both defect and bulk OPEs, and is consistent with available results coming from localization. The method is completely algorithmic and can be implemented to calculate correlators of arbitrary weight.}

\begin{document} 
	
\maketitle

\newpage

\flushbottom

\section{Introduction}
\label{sec:intro}

The analytic conformal bootstrap is a powerful tool that has seen significant progress in recent years. The basic proposal made in the original papers \cite{Fitzpatrick:2012yx,Komargodski:2012ek}, is that singular terms in a CFT correlator imply the existence of families of operators at large spin. This observation was later systematized and allowed the calculation of CFT data as a series expansion in inverse powers of the spin variable \cite{Alday:2016njk}. This line of thinking led to the Lorentzian inversion formula, that neatly captures the dependence of the CFT data as an analytic function of the spin variable \cite{Caron-Huot:2017vep}. This formula is a fully non-perturbative result valid for any CFT in any dimension. 

Apart from giving us an improved understanding of the structure of CFT, the inversion formula is also a powerful calculational tool. Its power lies on the fact that correlators can be reconstructed from a certain discontinuity, which is in general a simpler object. Inversion formula technology is particularly useful for planar theories at strong coupling, where the discontinuity is captured by only a finite number of conformal blocks, simplifying computations considerably. This approach was carried out successfully for four-point functions of half-BPS operators in $\mathcal{N}=4$ SYM \cite{Alday:2017vkk,Caron-Huot:2018kta}, and confirmed a previous conjecture made in Mellin space \cite{Rastelli:2016nze,Rastelli:2017udc}. 

Because the presence of a defect does not change the local physics, the main properties that enable the bootstrap for strong coupling correlators are still present if we add a line defect to the configuration. More precisely, in this paper we consider correlators between single trace half-BPS operators and a supersymmetric Wilson line. These are canonical examples of local and non-local operators in $\mathcal{N}=4$ SYM. One point-function of half-BPS operators in the presence of a line are fixed by kinematics and the overall coefficient can be calculated 
exactly using matrix-model techniques \cite{Semenoff:2001xp,Okuyama:2006jc,Giombi:2009ds,Billo:2018oog}. Less work has been done on two-point correlators in the presence of a line.
This is the first case in which correlators depend non-trivially on conformal invariants and therefore capture an infinite amount of CFT data. The only results on the literature so far are an explicit weak-coupling calculation \cite{Barrat:2020vch}, the extreme strong coupling limit where the correlator becomes the square of a one-point function \cite{Giombi:2012ep,Buchbinder:2012vr}, and special kinematical points where the correlator does not depend on invariants and can be calculated using localization \cite{Giombi:2012ep,Beccaria:2020ykg}. The goal of this paper is to calculate the full two-point correlator as a function of its cross-ratios, to the first non-trivial order at strong coupling, relying only on analytical bootstrap techniques.

The outline of the paper is as follows. In section \ref{sec:preliminaries} we review the basic kinematics of our two-point functions, including an interpretation of our bootstrap problem in terms of Witten diagrams. In section \ref{sec:22strong} we compute the two-point function for the stress-tensor multiplet, which is the half-BPS operator of weight $J=2$, using Lorentzian inversion technology, and present an elegant closed-form expression for the full correlator. In section \ref{sec:33and44} we generalize the analysis for operators of higher weight and present explicit solutions for the cases $J=3,4$ as a demonstration of our method. Possible future directions are discussed in section \ref{sec:conclusions}, while in appendix \ref{sec:appendix} we give the bulk and defect conformal blocks needed for the computations.

\section{Preliminaries}
\label{sec:preliminaries}

In this section we review the basic properties of the setup under study, 
before jumping into the technicalities of the computation in section \ref{sec:22strong}.

\subsection{Setup}
\label{subsec:CPO}

Consider $\Nm = 4$ SYM theory with gauge group $U(N)$ in the planar $N \to \infty$ limit.
The most important local operators in our discussion are single-trace chiral-primary operators:
\begin{equation}
\label{eq:bulkBPS}
\Op_J (x, u) 
:= (2\pi)^J \frac{2^{J/2}}{\sqrt{J \lambda^J}} 
   \tr \big( u \cdot \phi(x) \big)^J \,.
\end{equation}
Here $u$ is a six-component null polarization vector $u \cdot u = 0$, such that $\Om_J(x,u)$ transforms in a symmetric-traceless representation of the $R$-symmetry group.
These are protected operators that preserve half of the supercharges of the theory.
The other important observable in our analysis is the supersymmetric Wilson line (sometimes called the Maldacena-Wilson line):
\begin{equation}
\Wl := \frac{1}{N} \tr\, \Pm \exp \int_{-\infty}^{\infty} d\tau \bigl( i \tensor{\dot{x}}{^{\smash{\mu}}} \tensor{A}{_{\smash{\mu}}} + | \dot{x} | \theta\cdot \phi \bigr) \,.
\label{eq:MaldacenaWilsonLoop}
\end{equation}
This extended operator also preserves half of the supercharges, and has been studied extensively in the literature.
Here $\theta$ is a unit-normalized six-component vector $\theta \cdot \theta = 1$ that parametrizes a direction in $R$-symmetry space.

In this work we study correlators of local operators in the presence of the supersymmetric Wilson line:\footnote{The overall normalization is redundant in the present case because $\vev{\Wl} = 1$, but it would be important for circular Wilson loops. }
\begin{equation}
\vvev{\Op_{J_1}(x_{1}, u_1) \Op_{J_2}(x_{2}, u_2) \ldots }
:= \frac{1}{\vev{\Wl}} \vev{\Wl\, \Op_{J_1}(x_{1}, u_1)\, \Op_{J_2}(x_{2}, u_2) \ldots } \,.
\label{eq:correlator}
\end{equation}
Even though our configuration breaks some of the $PSU(2,2|4)$ symmetry of $\mathcal{N}=4$ SYM, these correlation functions are still restricted by the remaining defect (super)conformal symmetry.
It is well understood that in defect CFT one-point functions are kinematically fixed, see for example \cite{Billo:2016cpy}.
The simplest observables in which the coordinate dependence is dynamical and not fixed by symmetry  are two-point functions of bulk operators. The case $J_1 = J_2 = 2$ was studied recently at weak 't Hooft coupling \cite{Barrat:2020vch} using standard Feynman diagrams. Here we consider instead a perturbative expansion at large 't Hooft coupling:
\begin{align}
  \vvev{\Op_{J}(x_{1}, u_1) \Op_{J}(x_{2}, u_2) }
 \quad \text{as} \quad N \to \infty, \; \lambda = g^2 N \gg 1 \, .
\end{align}
To be more precise, our correlator admits a double expansion in powers of $\frac{\lambda}{N^2}$ and $\frac{1}{\sqrt{\lambda}}$ of the form
\begin{align}
\label{eq:perturb-structure}
  \vvev{\Op_{J} \Op_{J} }
  = \vvev{\Op_{J} \Op_{J} }^{(0)}
  + \frac{\lambda}{N^2} \left( 
    \vvev{\Op_{J} \Op_{J} }^{(1)}
  + \frac{1}{\sqrt{\lambda}} \vvev{\Op_{J} \Op_{J} }^{(2)}
  + O(\lambda^{-1})
  \right) 
  + \ldots \, ,
\end{align}
where $\ldots$ stands for terms starting at $\frac{\lambda^2}{N^4}$.
As we will discuss shortly, the first two terms in the expansion are somewhat trivial.
Our goal then is to use modern bootstrap methods to calculate the $\vvev{\Op_{J} \Op_{J} }^{(2)}$ contribution.
The above perturbative expansion has a natural interpretation in terms of the holographic dual of $\Nm = 4$ SYM, to which we now turn.

\subsection[Supergravity interpretation]{Supergravity interpretation}
\label{subsec:sugra}

Thanks to the $AdS$/CFT correspondence \cite{Maldacena:1997re,Witten:1998qj,Gubser:1998bc}, the strong coupling limit of $\Nm = 4$ SYM admits a description in terms of classical IIB supergravity on $AdS_5 \times S^5$.
Although we do not use this description to carry out our calculations, it is useful to understand the structure of perturbation theory.\footnote{We thank S. Giombi for useful correspondence regarding the holographic calculation.}

The dual of a supersymmetric Wilson loop is a string worldsheet extending inside $AdS_5$, whose boundary corresponds to the path of the loop \cite{JJ1,sjrey}.
Graphically
\begin{align}
 \begin{tikzpicture}[baseline={([yshift=-.55ex]current bounding box.center)},scale=0.67]
 \pgfmathsetmacro{\x}{sqrt(2)/2}
 \pgfmathsetmacro{\y}{sqrt(2)/2}
 \pgfmathsetmacro{\z}{0.5}
 \draw [thick] (0,0) circle [radius=1];
 \draw [thick, blue] (-\x,\y) to[out=-60,in=60] (-\x,-\y);
\end{tikzpicture}
= \vev{ \Wl }
= 1\, .
\end{align}
Here the black circle is the boundary of $AdS_5$, where the CFT lives, and the blue line corresponds to the string worldsheet.
The expectation value of the Wilson loop has been the subject of intense study \cite{Erickson:2000af,Drukker:2000rr,Pestun:2007rz}, but here we concentrate on the straight geometry.

We are interested on the interplay between the supersymmetric Wilson line and half-BPS single-trace operators $\Om_J$.
In the holographic description, $\Om_J$ are dual to certain KK modes arising from the compactification of the IIB action on $S^5$.
One of these modes can be sourced at the boundary of $AdS_5$, propagate through the bulk and be absorbed by the string worldsheet.
This process is dual to the one-point function of $\Om_J$ in the presence of the Wilson line. Graphically
\begin{align}
 \begin{tikzpicture}[baseline={([yshift=-.55ex]current bounding box.center)},scale=0.67]
 \pgfmathsetmacro{\x}{sqrt(2)/2}
 \pgfmathsetmacro{\y}{sqrt(2)/2}
 \pgfmathsetmacro{\z}{0.5}
 \draw [thick] (0,0) circle [radius=1];
 \draw [thick, blue] (-\x,\y) to[out=-60,in=60] (-\x,-\y);
 \draw (-\z, 0) -- (1, 0);
\end{tikzpicture}
= \vvev{\Om_J(x, u)}
= a_J \frac{(u \cdot \theta)^J}{(x^\bot)^J} \, .
\end{align}
The precise constant $a_J$ has been determined at strong 't Hooft coupling using holography \cite{Berenstein:1998ij, Giombi:2009ds}, and is consistent with the exact result coming from the matrix-model description.
The result is of order $a_J \sim O(\frac{\sqrt{\lambda}}{N})$, as can also be seen by simple power counting arguments.

The focus of the present work is on two-point functions at strong coupling.
The leading order contribution corresponds to a disconnected diagram, where the two operators in the boundary of $AdS_5$ interact through the bulk ``ignoring'' the string worldsheet:
\begin{align}
\label{eq:disc-correlator}
 \begin{tikzpicture}[baseline={([yshift=-.55ex]current bounding box.center)},scale=0.67]
 \pgfmathsetmacro{\x}{sqrt(2)/2}
 \pgfmathsetmacro{\y}{sqrt(2)/2}
 \pgfmathsetmacro{\z}{0.5}
 \draw [thick] (0,0) circle [radius=1];
 \draw [thick, blue] (-\x,+\y) to[out=-60,in=60] (-\x,-\y);
 \draw []            (+\x,+\y) to[out=-120,in=120] (+\x,-\y);
\end{tikzpicture}
= \vvev{ \Om_J(x_1, u_1) \Om_J(x_2, u_2) }_{\text{disc.}}
= \frac{(u_1 \cdot u_2)^J}{(x_{12}^2)^J} \, .
\end{align}
Following the usual convention in CFT, we normalize this diagram to unity.
This corresponds to $\vvev{ \Om_J \Om_J }^{(0)}$ in equation \eqref{eq:perturb-structure}.
The next contribution at strong coupling corresponds to a factorized diagram, where the operators do not interact in the bulk:
\begin{align}
\label{eq:mft-correlator}
 \begin{tikzpicture}[baseline={([yshift=-.55ex]current bounding box.center)},scale=0.67]
 \pgfmathsetmacro{\x}{sqrt(2)/2}
 \pgfmathsetmacro{\y}{sqrt(2)/2}
 \pgfmathsetmacro{\z}{0.5}
 \pgfmathsetmacro{\r}{0.52}
 \pgfmathsetmacro{\s}{0.3}
 \draw [thick] (0,0) circle [radius=1];
 \draw [thick, blue] (-\x,+\y) to[out=-60,in=60] (-\x,-\y);
 \draw []            (+\x,+\y) -- (-\r,+\s);
 \draw []            (+\x,-\y) -- (-\r,-\s);
\end{tikzpicture}
= \vvev{ \Om_J(x_1, u_1) } \vvev{ \Om_J(x_2, u_2) }
= a_J^2 \frac{(u_1 \cdot \theta)^J (u_2 \cdot \theta)^J}{(x_1^\bot x_2^\bot)^J} \, .
\end{align}
From the scaling of the one-point coefficients $a_J$, it is clear this diagram contributes at order $\frac{\lambda}{N^2}$, so it corresponds to the term $\vvev{ \Om_J \Om_J }^{(1)}$.
The first non-trivial correction to the two-point function contains an interaction vertex in the bulk \cite{Giombi:2012ep}.
In the exchanged line one sums over all KK modes that can couple to two $\Om_J$'s.
Schematically we have
\begin{align}
 \label{eq:witten-diagram-nontrivial}
 \begin{tikzpicture}[baseline={([yshift=-.55ex]current bounding box.center)},scale=0.67]
 \pgfmathsetmacro{\x}{sqrt(2)/2}
 \pgfmathsetmacro{\y}{sqrt(2)/2}
 \pgfmathsetmacro{\z}{0.5}
 \pgfmathsetmacro{\r}{-0.2}
 \draw [thick] (0,0) circle [radius=1];
 \draw [thick, blue] (-\x,+\y) to[out=-60,in=60] (-\x,-\y);
 \draw []            (+\x,+\y) -- (-\r,0);
 \draw []            (+\x,-\y) -- (-\r,0);
 \draw []            (-\r,  0) -- (-\z,0);
\end{tikzpicture} \, .
\end{align}
This is the diagram we calculate in the present work.\footnote{
In the case of holographic four-point functions, there exist contact Witten diagrams which are correctly reconstructed from the inversion formula \cite{Alday:2017vkk,Caron-Huot:2018kta}. If such diagrams are present in the defect case, our bootstrap result should capture them. We defer a more detailed study of the explicit holographic calculation for future work.}

Instead of an explicit calculation using the effective action in $AdS_5 \times S^5$, we will bootstrap the result using the bulk-to-defect inversion formula obtained in \cite{Lemos:2017vnx}.
The inversion formula reconstructs a correlator from its singular part, which is mathematically captured by a discontinuity:
\begin{align}
 \begin{tikzpicture}[baseline={([yshift=-.55ex]current bounding box.center)},scale=0.67]
 \pgfmathsetmacro{\x}{sqrt(2)/2}
 \pgfmathsetmacro{\y}{sqrt(2)/2}
 \pgfmathsetmacro{\z}{0.5}
 \pgfmathsetmacro{\r}{-0.2}
 \draw [thick] (0,0) circle [radius=1];
 \draw [thick, blue] (-\x,+\y) to[out=-60,in=60] (-\x,-\y);
 \draw []            (+\x,+\y) -- (-\r,0);
 \draw []            (+\x,-\y) -- (-\r,0);
 \draw []            (-\r,  0) -- (-\z,0);
\end{tikzpicture}
\;
\sim
\;
\int
\Disc \;
 \begin{tikzpicture}[baseline={([yshift=-.55ex]current bounding box.center)},scale=0.67]
 \pgfmathsetmacro{\x}{sqrt(2)/2}
 \pgfmathsetmacro{\y}{sqrt(2)/2}
 \pgfmathsetmacro{\z}{0.5}
 \pgfmathsetmacro{\r}{-0.2}
 \draw [thick] (0,0) circle [radius=1];
 \draw [thick, blue] (-\x,+\y) to[out=-60,in=60] (-\x,-\y);
 \draw []            (+\x,+\y) -- (-\r,0);
 \draw []            (+\x,-\y) -- (-\r,0);
 \draw []            (-\r,  0) -- (-\z,0);
\end{tikzpicture} \, .
\end{align}
The crucial property of holographic CFT's is that the discontinuity is dramatically simpler than the correlator.
In particular, we show below that the discontinuity receives corrections only from a finite number of single-trace operators exchanged in the bulk.
Each of these contributions is schematically a product of a tree-level one-point function and a three-point function:
\begin{align}
 \begin{tikzpicture}[baseline={([yshift=-.55ex]current bounding box.center)},scale=0.67]
 \pgfmathsetmacro{\x}{sqrt(2)/2}
 \pgfmathsetmacro{\y}{sqrt(2)/2}
 \pgfmathsetmacro{\z}{0.5}
 \pgfmathsetmacro{\r}{-0.2}
 \draw [thick] (0,0) circle [radius=1];
 \draw [thick, blue] (-\x,+\y) to[out=-60,in=60] (-\x,-\y);
 \draw []            (+\x,+\y) -- (-\r,0);
 \draw []            (+\x,-\y) -- (-\r,0);
 \draw []            (-\r,  0) -- (-\z,0);
\end{tikzpicture}
\;
\sim
\;
\sum_{\substack{\text{single} \\ \text{traces}}}
\int
 \begin{tikzpicture}[baseline={([yshift=-.55ex]current bounding box.center)},scale=0.67]
 \pgfmathsetmacro{\x}{sqrt(2)/2}
 \pgfmathsetmacro{\y}{sqrt(2)/2}
 \pgfmathsetmacro{\z}{0.5}
 \draw [thick] (0,0) circle [radius=1];
 \draw [thick, blue] (-\x,\y) to[out=-60,in=60] (-\x,-\y);
 \draw (-\z, 0) -- (1, 0);
\end{tikzpicture}
\times 
\begin{tikzpicture}[baseline={([yshift=-.55ex]current bounding box.center)},scale=0.67]
 \pgfmathsetmacro{\x}{1/2}
 \pgfmathsetmacro{\y}{sqrt(3)/2}
 \pgfmathsetmacro{\z}{0.5}
 \pgfmathsetmacro{\r}{-0.3}
 \pgfmathsetmacro{\q}{1.3}
 \draw [thick] (0,0) circle [radius=1];
 \draw []            (+\x,+\y) -- (0, 0);
 \draw []            (+\x,-\y) -- (0, 0);
 \draw []            (  0,  0) -- (-1, 0);
\end{tikzpicture} \, .
\end{align}
The one- and three-point functions are known from localization, so all is left is to compute a certain integral and sum over finitely many single-trace contributions.
In the rest of the paper, we translate this pictorial representation into a concrete bootstrap algorithm, that fully fixes two-point correlators with minimal external input.

\subsection{Superconformal kinematics}
\label{sec:kinematics}

The defect CFT associated with the Wilson line \eqref{eq:MaldacenaWilsonLoop} preserves an $OSp(4^*|4)$ subgroup of the full $PSU(2,2|4)$ symmetry.
From the spacetime perspective, the defect preserves an $SO(2,1) \subset SO(4,2)$ subgroup of the full conformal algebra.
Furthermore, the presence of a preferred direction $\theta$ preserves only $SO(5)_R \subset SO(6)_R$ of the $R$-symmetry.
The kinematics of non-supersymmetric defects have been thoroughly studied in \cite{Billo:2016cpy}.
Below we only give a brief review, highlighting the new features due to supersymmetry, such as superconformal Ward identities and superconformal blocks.

\subsubsection{Two-point function and cross-ratios}

A two-point function in a defect CFT depends on two spacetime cross-ratios, while in our setup there is an extra $R$-symmetry cross-ratio $\OR$:
\begin{equation}
\vvev{ \Op_J(x_1, u_1) \Op_J(x_2, u_2) } = 
\frac{(u_1 \cdot \theta)^J (u_2 \cdot \theta)^J}{|x_1^\bot|^J |x_2^\bot|^J} \Fm^{(J)} (z , \zb , \sigma)\,.
\label{eq:superconformalinvariance}
\end{equation}
For the spacetime part, we use the cross-ratios $z,\zb$ defined in \cite{Lemos:2017vnx}
\begin{align}
 \frac{z + \zb}{2 \sqrt{z \zb}}
 = \frac{x_1^\bot \cdot x_2^\bot}{|x_1^\bot| |x_2^\bot|} \, , \qquad
 \frac{(1-z)(1-\zb)}{\sqrt{z \zb}}
 = \frac{x_{12}^2}{|x_1^\bot| |x_2^\bot|} \, .
\end{align}
Geometrically $z,\zb$ are coordinates in a plane orthogonal to the defect. Indeed, placing the first operator at $x_1 = (1,0,0,0)$ and the second one in a $xy$-plane $x_2 = (x,y,0,0)$, then $z = x+i y$ and $\zb = x - iy$.
In Lorentzian signature, one would instead find that $z$ and $\zb$ are real and independent.
On the other hand, the $R$-symmetry cross-ratio is defined as
\begin{equation}
\OR := \frac{(u_1 \cdot u_2)}{(u_1 \cdot \theta)(u_2 \cdot \theta)} \,.
\label{eq:Rsymmetryinvariant}
\end{equation}
The correlator $\Fm^{(J)} (z , \zb , \sigma)$ is a polynomial of order $J$ in the $\sigma$ cross-ratio.
This reflects the number of different ways to contract $u_1, u_2$ with each other and with $\theta$, the polarization of the Wilson-line defect.
Before moving on, let us point out that one can also consider operators restricted to the line and study the corresponding $1d$ CFT, this configuration has been studied recently by a variety of means \cite{Giombi:2017cqn,Giombi:2018qox,Liendo:2018ukf,Ferrero:2021bsb,Grabner:2020nis,Cavaglia:2021bnz}\footnote{Similar setups have also been considered for $3d$ ABJM theories \cite{Bianchi:2017ozk,Bianchi:2018scb,Bianchi:2020hsz}.}. In this work we consider bulk operators outside the line, which probes the interplay between local physics and the defect.

\subsubsection{Structure of the correlator}

As discussed above, the leading contributions at strong coupling to $\vvev{\Om_J \Om_J}$ take the simple form \eqref{eq:disc-correlator} and \eqref{eq:mft-correlator}.
Comparing with the form of the two-point function \eqref{eq:superconformalinvariance} we find
\begin{align}
\label{eq:structure-LO}
\begin{split}
 \Fm^{(J)}(z, \zb, \OR)
 & =  \left( \frac{\OR \sqrt{z \zb}}{(1-z)(1-\zb)}\right)^J 
   + a_J^2
   + O\!\left(\frac{\sqrt{\lambda}}{N^2} \right) \, .
\end{split}
\end{align}
Our goal is to compute the correlation function at the next order $O(\frac{\sqrt{\lambda}}{N^2})$.
For simplicity we decompose the correlator at this order in powers of the $R$-symmetry cross-ratio $\OR$:\footnote{Note that the definition of $F^{(J)}_j$ in this decomposition is slightly different to the one used in \cite{Barrat:2020vch}: here the terms are ordered in terms of powers of $\OR$ and not $\Omega$. Furthermore, here we keep outside the definition the two leading contributions which are trivial.}
\begin{align}
\label{eq:structure-NLO}
\begin{split}
 \Fm^{(J)}(z, \zb, \OR) \big|_{O(\frac{\sqrt{\lambda}}{N^2})}
 & = \sum_{j=0}^J \OR^j F^{(J)}_{J-j}(z, \zb) \, .
\end{split}
\end{align}
Below we use bootstrap methods to reconstruct the functions $F_j^{(J)}(z, \zb)$,
leading to the final results in equations \eqref{eq:completion22}, \eqref{eq:full-corr-33} and \eqref{eq:full-corr-44}.

\subsubsection{Superconformal Ward identities}

An important property of our correlator is that it satisfies \textit{superconformal Ward identities}. 
These were studied in detail for half-BPS boundaries in \cite{Liendo:2016ymz}, where it was observed that with a suitable identification of the cross-ratios, they also apply to the half-BPS line defect.\footnote{To be precise, $z,\zb$ are mapped to the boundary $R$-symmetry cross-ratios $w_1$, $w_2$, while $\sigma$ is mapped to the boundary spacetime cross-ratio $-2\xi$ (this factor $-2$ ensures the correctness of this map).}
In our conventions the Ward identities read
\begin{equation}
\left. \left( \partial_z + \frac{1}{2} \partial_\omega \right) 
\Fm^{(J)} (z,\zb,\OR) \right|_{z = \omega} = 0\,, \quad 
\left. \left( \partial_{\zb} + \frac{1}{2} \partial_\omega \right) 
\Fm^{(J)} (z,\zb,\OR) \right|_{\zb = \omega} = 0\, ,
\label{eq:WI}
\end{equation}
where the natural variable to use is $\omega$, defined by
\begin{equation}
\OR = -\frac{(1-\omega)^2}{2 \omega}\,.
\label{eq:invariant}
\end{equation}
It is not hard to check that the two leading terms \eqref{eq:structure-LO} satisfy these equations.
Note that the above Ward identities take a form very similar to other setups in the literature \cite{Dolan:2004mu,Nirschl:2004pa,Liendo:2015cgi}.
In the case of $\Nm=4$ SYM without defects, the Ward identities admit a simple closed-form solution in which the different $R$-symmetry channels are related by algebraic relations. This drastically simplifies the analysis because in the end one can just work with the independent channels. Sadly, the defect setup of this paper is closer to the Ward identities in three dimensions, where such an algebraic relation does no exist \cite{Dolan:2004mu}. It is possible however to relate the different $R$-symmetry channels by the action of a non-local operator. This is obviously more cumbersome, but one can still implement it in order to focus only on the independent portions of the correlator \cite{Chester:2014fya}. In this paper we work explicitly with all the $R$-symmetry channels, however finding a better parameterization for our correlators is an interesting problem that should be studied in more detail. 

\subsubsection{Crossing equation}

The last important ingredient for our calculation are the defect CFT crossing equations.
Here we only summarize them, and refer the reader to \cite{Billo:2016cpy} for further details.
In any CFT it is possible to fuse two bulk local operators as a sum of bulk operators.
This \emph{bulk OPE} is denoted schematically as
$\Op_{J} \Op_{J} \sim \sum_{\Op} \lambda_{JJ\Om} \Om$.
On the other hand, the \emph{defect OPE} expands a bulk operator as a sum of defect operators $\Op_J \sim \sum_{\Oh} b_{J\Oh} \Oh$.
Note that for a bulk one-point function $\vvev{ \Om } \sim a_\Om$, the defect OPE implies $a_{\Om} = b_{\Om \hat{\mathds 1}}$ where $\hat{\mathds 1}$ is the defect identity operator.
These two expansions can be inserted in a two-point function, resulting in a \emph{crossing equation}
\begin{equation}
 \Fm^{(J)}(z, \zb, \OR)
 = \left( \frac{\sqrt{z \zb} \, \OR}{(1-z)(1-\zb)}\right)^J 
   \sum_{\Op} \lambda_{JJ\Om} a_\Om \Gm^{(J)}_\Om (z,\zb,\OR) 
 = \sum_{\Oh} b^2_{J\Oh} \Gh^{(J)}_{\Oh} (z,\zb,\OR)\,.
\label{eq:crossing}
\end{equation}
It is important to keep in mind that operators in superconformal theories belong to superconformal multiplets.
For example, in the crossing equation, $\Op$ refers to the superconformal primary operator of a $PSU(2,2|4)$ representation, which is labeled by the dimension $\Delta$, the spin $\ell$, and an $SO(6)_R$ label $K$.
Similarly, $\Oh$ is the superconformal primary of an $OSp(4^*|4)$ representation labeled by the defect dimension $\Dh$, the transverse-spin $s$, and an $SO(5)_R$ label $\hat K$.\footnote{Detailed analysis of the representation theory of these supergroups can be found in the literature \cite{Dolan:2002zh,Cordova:2016emh,Agmon:2020pde}.}
Supersymmetry fixes the contributions of superdescendant operators, so the expansion \eqref{eq:crossing} is organized in \emph{superconformal blocks}.
These are linear combination of non-supersymmetric conformal blocks:
\begin{align}
\begin{split}
 & \Gm^{(J)}_\Om (z,\zb,\OR) 
 = \sum_{\Delta,\ell,K} c_{\Delta,\ell,K}^{(J)} h_K (\OR) f_{\Delta,\ell} (z,\zb)\,, \\
 & \Gh^{(J)}_{\Oh} (z,\zb,\OR) 
 = \sum_{\Dh,s,\hat K} c_{\Dh,s,\hat K}^{(J)} \hh_{\hat K} (\OR) \fh_{\Dh,s} (z,\zb)\,.
\end{split}
\end{align}
The sums range over all operators in the supermultiplets of $\Om$/$\Oh$ that can appear in the bulk/defect OPE of $\Om_J$. The function $f_{\Delta,\ell}(z,\zb)$ is a bulk-channel conformal block, see \eqref{eq:bulkstblock} for a useful series representation. 
Similarly, $\hat f_{\Dh,s}(z,\zb)$ is a defect-channel conformal block given in \eqref{eq:defect-block}.
Finally, $h_K (\OR)$ and $\hh_{\hat K}(\OR)$ are bulk and defect $R$-symmetry blocks respectively.
They are polynomials in $\OR$ with a simple hypergeometric closed form expression \eqref{eq:R-sym-block}.
The precise relative coefficients $c_{\Delta,\ell,K}^{(J)}$ and  $c_{\Dh,s,\hat K}^{(J)}$ have been presented in the appendix of \cite{Barrat:2020vch}.

\subsection{Topological subsector}
\label{sec:top-sector}

An important property of our correlators is that they contain a topological subsector. In terms of the cross-ratios, the topological subsector can be obtained by setting $z=\bar{z}=\omega$. This projects out all the non-protected operators from the bulk and defect OPE and only half-BPS protected operators remain. This topological subsector goes beyond our two-point correlators, and it actually describes a closed-subsector of the operator spectrum of the theory. 
The CFT data of the topological subsector can be obtained by solving Gaussian multi-matrix models \cite{Giombi:2009ds,Giombi:2012ep,Giombi:2018qox,Giombi:2018hsx}.
Some of the explicit results that we present below are necessary to fix overall coefficients in our bootstrap analysis, while others provide non-trivial consistency checks of our calculation.

We start by looking at bulk single-trace half-BPS operators $\Om_J$ defined in \eqref{eq:bulkBPS}.
We follow the usual normalization conventions such that their two-point function is unit normalized, and the dynamical information is captured by three-point functions.
In our conventions
\begin{align}
\begin{split}
 \label{eq:norm-top-2pt-3pt}
  \vev{\Op_{J}(x_1, u_1) \Op_{J}(x_2, u_2)  } 
  &= \frac{(u_1 \cdot u_2)^J}{(x_{12}^2)^J} \, , \\
  \vev{\Op_{J_1}(x_1, u_1) \Op_{J_2}(x_2, u_2) \Op_{J_3}(x_3, u_3) } 
  &= \lambda_{J_1 J_2 J_3} \frac{(u_1 \cdot u_2)^{J_{123}} (u_2 \cdot u_3)^{J_{231}} (u_3 \cdot u_1)^{J_{312}}}{|x_{12}|^{2J_{123}} |x_{23}|^{2J_{231}} |x_{31}|^{2J_{312}}}\,,
\end{split}
\end{align}
with $J_{ijk} := (J_i + J_j - J_k)/2$. The OPE coefficients $\lambda_{J_1 J_2 J_3}$ were originally computed in \cite{Lee:1998bxa}, and they are independent of the coupling:
\begin{align}
\label{eq:single-trace-localization-1}
\begin{split}
 \lambda_{J_1 J_2 J_3}
 = \frac{\sqrt{J_1 J_2 J_3}}{N} \, . \\
\end{split}
\end{align}
As already discussed, one-point functions of single-trace operators in the presence of the supersymmetric Wilson line are kinematically fixed:
\begin{equation}
\label{eq:normalization-1pt}
\vvev{\Op_J (x, u)} = a_J \frac{(u \cdot \theta)^J}{|x^\bot|^J}\,.
\end{equation}
Since the normalization of $\Om_J$ is determined by the two point function, the coefficient $a_J$ contains dynamical information about the defect CFT.\footnote{For the particular case $J=2$ the coefficient $a_2$ is proportional to the Bremsstrahlung function \cite{Correa:2012at}.}
The precise value can be obtained from a perturbative calculation \cite{Semenoff:2001xp} or from a matrix-model calculation \cite{Okuyama:2006jc,Giombi:2009ds,Billo:2018oog}
\begin{align}
\label{eq:single-trace-localization}
\begin{split}
 & a_J 
 = \frac{\sqrt{\lambda  J}}{2^{J/2+1} N}
   \frac{I_J(\sqrt{\lambda })}{I_1(\sqrt{\lambda })} 
 \overset{\lambda \gg 1}{=} \frac{\sqrt{\lambda J}}{2^{J/2+1} N} \left(
      1
    - \frac{(J+1)(J-1)}{2 \sqrt{\lambda }}
    + O(\lambda^{-1}) \right)\, .
\end{split}
\end{align}

Note that it is also possible to construct half-BPS multi-trace operators. 
For example, the OPE between the two single-trace operators $\Om_J \times \Om_J$ must contain the following \textit{double-trace} operator:
\begin{equation}
\Op_{(J,J)} (x, u) 
 := (2 \pi)^{2J} \frac{2^{J-1/2}}{J \lambda^J} \, \tr \big(u \cdot \phi(x) \big)^J \tr\big(u \cdot \phi(x) \big)^J\,.
\label{eq:doubletracebulk}
\end{equation}
The prefactor is chosen to have a unit normalized two-point function.
By using the localization techniques of \cite{Giombi:2009ds,Billo:2018oog,Beccaria:2020ykg}, we derived the one-point function of $\Op_{(J,J)}$ and its three-point function with single-trace operators.
Their product reads
\begin{align}
\label{eq:double-trace-localization}
 a_{(J,J)} \lambda_{JJ(J,J)} 
 = \frac{J \lambda }{2^{J+2} N^2}
   \frac{I_{2 J-1}(\sqrt{\lambda})}{I_1(\sqrt{\lambda})} 
 \overset{\lambda \gg 1}{=} \frac{J \lambda}{2^{J+2} N^2} \left(
    1 
   - \frac{2 J (J-1)}{\sqrt{\lambda }}
   + O(\lambda^{-1})
 \right)\, .
\end{align}
This is all the information we need about bulk protected operators.

The topological sector also captures information involving protected operators localized on the defect.
There is a large degeneracy of protected defect operators which has been discussed in \cite{Giombi:2018hsx}.
For our purposes, there are two defect operators that play an important role.
On the one hand, there are defect operators inserted inside the path-ordering of the Wilson line, which we write as
\begin{align}
 \hat \Om_{\hat K}(\tau, \hat u) 
 = \Wm \Big[ \big(\hat u \cdot \phi(\tau) \big)^{\hat K} \, \Big] \, .
\end{align}
These operators are in symmetric traceless representations of the $SO(5)_R$ symmetry preserved by the defect.
Therefore, $\hat u$ is a  null polarization vector orthogonal to the Wilson line polarization $\hat u \cdot \theta = \hat u \cdot \hat u = 0$.
On the other hand, it is possible to define a similar operator which lives outside the path-ordering
\begin{align}
 \hat \Om_{(\hat K)}(\tau, \hat u) 
 = \Wm \tr \big( \hat u \cdot \phi(\tau) \big)^{\hat K} \, .
 \label{eq:defectdoubletrace}
\end{align}
For both of these operators, one should choose a normalization such that their defect two-point functions are unit normalized.
The non-trivial dynamical information is then encoded in the bulk-defect two-point function:
\begin{align}
\begin{split}
 \vvev{\hat \Op_{\hat K} (\tau_1, \uh_1) \hat \Op_{\hat K} (\tau_2, \uh_2)} &= \frac{(\hat u_1 \cdot \hat u_2)^{\hat K}}{\tau_{12}^{2\hat K}} \, , \\
 \vvev{\Op_{J} (x_1, u_1) \Oh_{\hat K} (\tau_2, \hat u_2)} &= b_{J\hat K} \frac{(u_1 \cdot \uh_2)^{\hat K} (u_1 \cdot \theta)^{J-\hat K}}{((x_1^\bot)^2 + \tau_{12}^2)^{\hat K} |x_1^\bot|^{J-\hat K}}\, .
\label{eq:bulkdefecttwopt} 
\end{split}
\end{align}
For operators of the first type $\hat \Om_{\hat K}$, the bulk-defect coefficient has been calculated at strong coupling in the planar limit \cite{Giombi:2018hsx}
\begin{align}
\label{eq:bulk-defect-localization}
 b_{J\hat K} =
 \frac{\lambda ^{\frac{2-\hat K}{4}}}{N}
 \frac{2^{\frac{3\hat K-J-2}{2}} \sqrt{J}}{\sqrt{\hat K!}}
 \frac{\Gamma(\frac{J+\hat K+1}{2})}{\Gamma(\frac{J-\hat K+1}{2})}
 \left(1 + \frac{4-4 J^2+5 \hat K + \hat K^2}{8 \sqrt{\lambda }} + O(\lambda^{-1}) \right) \, .
\end{align}
On the other hand, the coefficients $b_{J(\hat K)}$ have not appeared in the literature.
An interesting outcome of our analysis is a prediction for the value of $b_{J(\hat J)}$.
However, let us stress that our correlators also contain information about infinitely many non-protected operators that are not captured by the topological subsector.

Note that the crossing equation given in \eqref{eq:crossing} truncate on both sides in the topological sector \cite{Liendo:2016ymz}. This is known as \textit{microbootstrap}, and in some cases the system of equations can be solved exactly \cite{Barrat:2021un}.

\section{\texorpdfstring{$\vvev{\Op_2 \Op_2}$}{<< O2 O2>>} at strong coupling}
\label{sec:22strong}

In this section we compute the correlator $\vvev{\Op_2 \Op_2}$ in the strong coupling limit.
As explained in section \ref{subsec:sugra}, the two leading contributions are somewhat trivial and take the form given in equations \eqref{eq:disc-correlator} and \eqref{eq:mft-correlator}.
As in equation \eqref{eq:structure-NLO}, we decompose the next correction into three $R$-symmetry channels:
\begin{equation}
\Fm^{(2)} (z,\zb,\OR)\big|_{O(\frac{\sqrt{\lambda}}{N^2})} 
= \OR^2 F_0 (z,\zb) + \OR F_1 (z,\zb) + F_2 (z,\zb)\,.
\label{eq:Rsymm22}
\end{equation}
For compactness, here and for the rest of this section we drop the superscripts in the functions $F^{(J)}_j(z,\zb)$.
In what follows we derive this correlator using the Lorentzian inversion formula presented in \cite{Lemos:2017vnx}.

\subsection{Lorentzian inversion formula}
\label{subsec:LIF}

The idea of the Lorentzian inversion formula is that the \textit{discontinuity} of the correlator is sufficient to extract the full defect CFT data, which in turn can be used for reconstructing the full correlator. 
For now we consider general single-trace operators $\Om_J(x,u)$, and later we focus on the $J=2$ case.
For a codimension-three defect, such as a Wilson line in four dimensions, the inversion formula reads
\begin{align}
\begin{split}
 b_j (\hat \Delta, s) 
 &= \int_0^1 \frac{dz}{2z} z^{-(\Dh-s)/2} \int_1^{1/z} \frac{d\zb}{2\pi i}
   (1 - z \bar z) (\bar z - z) \bar z^{-(\hat \Delta + s)/2 - 2} \\
 & \quad \times 
   {}_2F_1 \left( \frac{1}{2}, 1+s, \frac{3}{2}+s; \frac{z}{\zb} \right)
   {}_2F_1 \left( \frac{1}{2}, 1-\Dh, \frac{3}{2}-\Dh, z \zb \right)
   \text{Disc} \, F_j(z, \zb) \, ,
\end{split}
\label{eq:inversion}
\end{align}
where the discontinuity is computed around the branch cut in $\zb \in [1, \infty)$:
\begin{align}
\label{eq:disc-def}
 \text{Disc} \, F_j(z, \zb)
 = F_j(z, \zb + i\epsilon) - F_j(z, \zb - i\epsilon), \qquad \zb \ge 1.
\end{align}
The inversion formula is \textit{bosonic} and thus it should be applied to each $R$-symmetry channel $F_j(z,\zb)$ independently.
The defect conformal dimensions are encoded in the poles of $b_j(\Dh,s)$, while the residues are OPE coefficients:
\begin{equation}
b_j(\Dh,s)
= - \sum_{n \geq 0} \frac{(b_j^2)_{n,s}}{\Dh - (J + s + 2n + \gamma_{n,s})} \,.
\end{equation}
The coefficients $(b_j)_{n,s}$ capture the normalization of the
bulk-defect two-point function $\vvev{\Om_J \hat \Om_{n,s}}$, where the exchanged operators $\hat\Om_{n,s}$ have dimension $\Dh_{n,s} = J+s+2n+\gamma_{n,s}$ and transverse-spin $s$.\footnote{In general the defect spectrum contains degeneracies, in which case the CFT data has to be understood as a sum over degenerate operators.} 
In the case where (small) anomalous dimensions are relevant, second-order poles are also present in the formula when Taylor expanding:
\begin{align}
b_j(\Dh,s)
& = - \sum_{n \geq 0} \left( \frac{(b_j^2)_{n,s}}{\Dh - (J + s + 2n)} + \frac{(b_j^2 \gamma_j)_{n,s}}{(\Dh - (J + s + 2n))^2} + \ldots \right)\,.
\label{eq:expb}
\end{align}
Once the OPE coefficients and the anomalous dimensions have been obtained, the correlator can be expanded in defect spacetime blocks: 
\begin{equation}
\tilde{F}_j (z,\zb) 
= \sum_{s=0}^\infty \sum_{n=0}^\infty \left( 
  (b_j^2)_{n,s} \hat f_{J + s + 2n, s}(z, \zb) 
+ (b_j^2 \gamma_j)_{n,s} \partial_{\Dh} \hat f_{J +s + 2n, s}(z, \zb) \right)\,.
\label{eq:Ftilde}
\end{equation}
Note that we have introduced the notation $\tilde{F}_j$ instead of $F_j$, since the inversion formula might miss contributions from low spins $s < s_*$ \cite{Lemos:2017vnx}. 
In that case, we must add extra terms with spins $s = 0, 1, \ldots, s_*$ to $\tilde{F}_j$ in order to recover the full correlator $F_j$. 
This procedure will be described in detail in section \ref{subsec:supersym22}.
Except for these subtleties, equations \eqref{eq:inversion}-\eqref{eq:Ftilde} reconstruct the function $F_j$ using only information in its discontinuity $\Disc F_j$.

\subsection{Computation of the discontinuity}
\label{subsec:OPEdisc}

The first step in order to apply the inversion formula \eqref{eq:inversion} is to compute the discontinuity.
As we now show, the discontinuity can be computed even though the full correlator is not known.

The idea is to expand the correlation functions in the bulk channel.
The conformal blocks in \eqref{eq:bulkstblock} can be written as $f_{\Delta,\ell}(z, \zb) = [(1-z)(1-\zb)]^{(\Delta-\ell)/2} \tilde f_{\Delta,\ell}(z,\zb)$, where the function $\tilde f_{\Delta,\ell}(z,\zb)$ has an expansion around $z,\zb = 1$ in positive integer powers.
As a result, only the prefactor can have non-vanishing discontinuity. 
Therefore, the contribution of a single bulk operator $\Om$ to the discontinuity is
\begin{align}
\begin{split}
 \Disc F_j(z, \zb) \big|_\Om
 & \propto (z \zb)^{J/2}
 (1-z)^{\frac{\Delta - (2J + \ell)}{2}}
 \tilde f_{\Delta,\ell}(z, \zb)
 \Disc \left[ (1-\zb)^{\frac{\Delta - (2J + \ell)}{2}} \right] \, .
\end{split}
\end{align}
There are two situations when this discontinuity does not vanish:
\begin{enumerate}
 \item If $\Delta$ is non-integer. This corresponds to $\Om$ having an anomalous dimension correcting its tree-level dimension $\Delta = 2J + \ell + 2n + \gamma$.
 \item If $\Delta$ is integer but $\Delta < 2J + \ell$. This corresponds to $\Om$ being a protected single-trace operator, whose dimension is below the double-trace threshold.
 Note that even though $\Disc (1-\zb)^{-n}$ naively vanishes, for $n>0$ the singularity at $\zb=1$ gives a finite contribution to the inversion formula.\footnote{This will be proved concretely by deriving equation \eqref{eq:basic_inversion}.}
\end{enumerate}

In our setup the discontinuity only receives contributions of the second type. This claim can be proved by studying the superconformal bulk OPE in detail.
From now on we focus on the $J=2$ case.
It was shown in \cite{Liendo:2016ymz} that in the presence of the line defect the OPE $\Op_2 \times \Op_2$ truncates in the following way:
\begin{equation}
\Op_2 \times \Op_2 \to \mathds{1} + \Bm_{[0,2,0]} + \Bm_{[0,4,0]} + \sum_\ell \Cm_{[0,2,0],\ell} + \sum_{\Delta, \ell} \Am_{[0,0,0],\ell}^\Delta\,.
\label{eq:bulkOPE22}
\end{equation}
These representations correspond to the operators acquiring a non-vanishing one-point function in the presence of a half-BPS line defect (like the supersymmetric Wilson line we study in this work). 
The operators in the $\Bm_{[0,2,0]}$ multiplet have integer dimension $\Delta < 4+\ell$, so they have non-vanishing discontinuity.
The operators in the $\Bm_{[0,4,0]}$, $\Cm_{[0,2,0],\ell}$ multiplets have integer dimension $\Delta \ge 4+\ell$, so they cannot contribute to the discontinuity.
Only the unprotected multiplets $\Am_{[0,0,0],\ell}^\Delta$ can have anomalous dimensions.
The scaling dimensions of these multiplets have the following schematic structure \cite{Goncalves:2014ffa}
\begin{equation}
\Delta = 2J + 2n + \ell + \frac{1}{N^2} \left( a + \frac{b}{\lambda^{3/2}} + \ldots \right) + O(N^{-4})\,,
\end{equation}
which means anomalous dimensions do not contribute at the order in perturbation theory we are working.
This implies all the operators in the bulk OPE \eqref{eq:bulkOPE22} do not have anomalous dimensions at this order, so the correlator must admit an expansion in integer powers in the limit $z, \zb \to 1$.

The main consequence of these observations is that only the superblock $\Gm_{[0,2,0]}$ corresponding to $\Bm_{[0,2,0]}$ has non-vanishing discontinuity:
\begin{equation}
\left. \text{Disc}\, \Fm^{(2)} (z,\zb,\OR) \right|_{O(\frac{\sqrt{\lambda}}{N^2})} = \lambda_{222} a_2\, \text{Disc} \left( \frac{\sqrt{z\zb} \, \sigma}{(1-z)(1-\zb)} \right)^2\Gm_{[0,2,0]} (z,\zb,\OR)\,.
\end{equation}

In the rest of this section we reconstruct the full correlator from the single superblock $\Gm_{[0,2,0]}$.

\subsection{Inversion of \texorpdfstring{$\Bm_{[0,2,0]}$}{B[020]}}
\label{subsec:inv020}

We now invert the superblock $\Gm_{[0,2,0]}$ in order to extract the defect CFT data, and by resumming the defect expansions, we obtain the correlators given in \eqref{eq:tildeF0}, \eqref{eq:tildeF1} and \eqref{eq:tildeF2}.

The superblocks $\Gm_{[0,K,0]}$ are known and given in equation \eqref{eq:protected-superblock}. For $K=2$ they take the form
\begin{equation}
\Gm_{[0,2,0]} (z,\zb,\OR) = h_2(\OR) f_{2,0}(z,\zb) + \frac{1}{180} h_0 (\OR) f_{4,2} (z,\zb)\,,
\label{eq:Gm020}
\end{equation}
where $h_K(\sigma)$ and $f_{\Delta,\ell}(z,\zb)$ correspond respectively to $R$-symmetry and spacetime conformal blocks, which can be found in appendix \ref{sec:appendix}. We see that we need to invert two bosonic blocks, namely a scalar block $f_{2,0}$ as well as the stress-tensor block $f_{4,2}$.
Using the definition \eqref{eq:R-sym-block} for the $R$-symmetry blocks, we can extract the discontinuities in the three channels:
\begin{align}
\begin{split}
\Disc F_0(z,\zb) &= - \lambda_{222} a_2 \Disc \frac{z \zb}{180 (1-z)^2 (1-\zb)^2} (30 f_{2,0}(z,\zb)-f_{4,2}(z,\zb))\,, \\
\Disc F_1 (z,\zb) &= \lambda_{222} a_2 \Disc \frac{z \zb}{(1-z)^2 (1-\zb)^2} f_{2,0} (z,\zb)\,, \\
\Disc F_2 (z,\zb) &= 0\,.
\end{split}
\label{eq:discsRsym}
\end{align}
It is convenient to express the blocks using the following variable:
\begin{equation}
\yb := \frac{1-\zb}{\sqrt{\zb}}\,.
\label{eq:ybdef}
\end{equation}
As discussed in the previous section, only negative powers of $\yb$ are relevant for the discontinuity.
Since a factor $\yb^{-2}$ comes from the prefactor in \eqref{eq:discsRsym}, we have to expand the bulk blocks to order $O(\yb)$ as $\yb \to 0$.
Using the methods in appendix \ref{sec:sing-blocks} we find
\begin{align}
\begin{split}
f_{2, 0}(z, \zb) 
& = -\yb \log z + O(\yb^3) \,, \\
f_{4, 2}(z, \zb) 
& = \yb \left(\frac{90 (z+1)}{z-1}-\frac{30 \left( z^2+4 z+1 \right) \log z}{(z-1)^2}\right)
+ O(\yb^3)\,.
\end{split}
\label{eq:expbosblocks}
\end{align}

Let us now invert an arbitrary power $\yb^{-p}$.
In this section we only need the $p = 1$ case, but the case $p \ge 2$ is relevant for section \ref{sec:33and44}.
For arbitrary powers of $\yb$ the discontinuity \eqref{eq:disc-def} results in:
\begin{equation}
\label{eq:disc-gen-p}
\Disc \yb^{-p} = 2 i \sin (p \pi) (- \yb)^{-p}\, .
\end{equation}
In principle now we should compute the inversion integral \eqref{eq:inversion}.
In practice this is too hard, so we expand the integrand as $z \to 0$.\footnote{As a side effect, the expansion as $z \to 0$ makes the inversion integral convergent order by order in $z$.}
Each new power of $z$ will give information of a new defect family with higher transverse-twist $\Dh - s$.
In the $z \to 0$ limit the integral over $\zb$ is standard and gives the following result:
\begin{align}
\label{eq:basic_inversion}
B_p(\beta) = 2i \sin ( \pi p) \int_1^\infty \frac{d\zb}{2\pi i} \zb^{-\beta/2 - 1} (-\yb)^{-p} = \frac{\Gamma \left(\frac{\beta+p}{2}\right)}{\Gamma(p) \Gamma \left(\frac{\beta-p+2}{2}\right)} \, .
\end{align}
Although for integer $p > 0$ the discontinuity \eqref{eq:disc-gen-p} naively vanishes, note that the final result is perfectly finite.  This is expected, because the correlator is singular at $\zb \to 1$, and the inversion formula reconstructs the CFT data from this singularity.
Note that for $p=1$ then $B_1(\beta) = 1$, which simplifies our calculations below.

In general the integral over $z$ has the following structure:
\begin{align}
\begin{split}
 b(\Dh,s) &= \sum_{n \geq 0} \int_0^1 \frac{dz}{2z} z^{-\frac{\Dh - (2 + s + 2n)}{2}} \left[ b^{(0,n)} (\Dh, s) + b^{(1,n)} (\Dh,s) \log z \right] \\
& = - \sum_{n \geq 0} \left( \frac{b^{(0,n)} (\Dh,s) + 2 \partial_{\Dh} b^{(1,n)} (\Dh,s)}{\Dh - (2 + s + 2n)} + \frac{2 b^{(1,n)} (\Dh,s)}{(\Dh - (2 + s + 2n))^2} + \ldots \right)_{\Dh \mathrlap{= 2 + s + 2n} } \,.
\end{split}
\label{eq:structureintegrals}
\end{align}
This has to be compared to equation \eqref{eq:expb} in order to obtain the OPE coefficients as well as the product of anomalous dimensions with tree-level OPE coefficients.
The presence of logs in equation \eqref{eq:expbosblocks} reveals that the scaling dimensions of the defect operators receive anomalous corrections at this order. 

In principle the results above are sufficient for extracting the defect CFT data in an algorithmic way and for resumming the correlator using equation \eqref{eq:Ftilde}. 
However, for the $p = 1$ case we can derive a closed-form formula for the defect CFT data corresponding to the bosonic blocks of equation \eqref{eq:Gm020}.
In the following we denote by $b_{f_{\Delta,\ell}} (\hat{\Delta},s)$ the result of the inversion formula performed for individual bosonic spacetime blocks $f_{\Delta,\ell}$. We begin with the scalar block $f_{2,0}$. Using the inversion formula as well as the integrals \eqref{eq:basic_inversion} and \eqref{eq:structureintegrals} we find
\begin{align}
b_{f_{2,0}}(\hat{\Delta},s)
&= - \sum_{j,k \geq 0} \frac{(s+1)_j (1/2)_j}{j! (s+3/2)_j} \frac{(1-\Dh)_k (1/2)_k}{k! (3/2-\Dh)_k} \int_0^1 \frac{dz}{2z} z^{-\frac{\hat{\Delta}-s-2}{2}} z^{j+k} \log z \notag \\
&= - \sum_{j,k \geq 0} \frac{(s+1)_j (1/2)_j}{j! (s+3/2)_j} \frac{(1-\Dh)_k (1/2)_k}{k! (3/2-\Dh)_k} \frac{- 2}{( \hat{\Delta} -s -2(j+k+1) )^2}\,,
\end{align}
where we have expanded the hypergeometric functions of equation \eqref{eq:inversion}. There are only second-order poles present, thus only the coefficients $b^{(1,n)}$ are non-trivial in equation \eqref{eq:structureintegrals}.
The infinite sum can be obtained in closed form
\begin{align}
\begin{split}
b_{f_{2,0}}^{(0,n)} (\Dh,s) &= 0\,, \\
b_{f_{2,0}}^{(1,n)} (\Dh,s) &= \Cm^{(n)} (\Dh,s)\,,
\end{split}
\end{align}
where
\begin{align}
 \Cm^{(n)}(\Dh, s)
 & =
  - \frac{\Gamma \left(n+\frac{1}{2}\right) (\Dh -n)_n}
      {n! \sqrt{\pi} \left(\Dh - n -\frac{1}{2}\right)_n}
 {}_4F_3\left( {\begin{array}{*{20}{c}}
    {\frac{1}{2},-n,s+1,\Dh - n -\frac{1}{2} } \\
    {\frac{1}{2}-n,s+\frac{3}{2},\Dh -n}
    \end{array}; 1} \right)\,.
\end{align}
The calculation of the stress-tensor block proceeds in an analogous way. Using the integrals given above we obtain
\begin{align}
\begin{split}
b_{f_{4,2}}^{(0,n)} (\Dh,s) &= 90 \Cm^{(n)} (\Dh,s) + 180 \sum_{m=1}^n \Cm^{(n-m)} (\Dh,s)\,, \\
b_{f_{4,2}}^{(1,n)} (\Dh,s) &= 30 \Cm^{(n)} (\Dh,s) + 180 \sum_{m=1}^n m\, \Cm^{(n-m)} (\Dh,s)\,.
\end{split}
\end{align}

Putting everything together, with the relative coefficients given by equation \eqref{eq:discsRsym}, the bosonic CFT data for $\tilde F_0(z,\zb)$ reads
\begin{align}
(b_0^2)_{n,s} &= \frac{1}{180} \lambda_{222} a_2 \left( b_{f_{4,2}}^{(0,n)} (\Dh,s) - 60\, \partial_{\Dh} b_{f_{2,0}}^{(1,n)} (\Dh,s) + 2 \partial_{\Dh} b_{f_{4,2}}^{(1,n)} (\Dh,s) \right)_{\Dh = J + s + 2n}\,, \notag \\
( b_0^2 \gamma_0)_{n,s} 
&= - \frac{1}{90} \lambda_{222} a_2 \left( 30\, b_{f_{2,0}}^{(1,n)} (\Dh,s) - b_{f_{4,2}}^{(1,n)} (\Dh,s) \right)_{\Dh = J + s + 2n}\,.
\end{align}
This can be used for resumming the correlator as in equation \eqref{eq:Ftilde}.
The result takes a very simple form:
\begin{equation}
\tilde{F}_0 (z, \zb) = - \lambda_{222} a_2 \frac{z \zb}{2(1-z)(1-\zb)} \left[ \frac{1+z \zb}{(1- z\zb)^2} + \frac{2 z \zb \log z\zb }{(1-z\zb)^3} \right]\,.
\label{eq:tildeF0}
\end{equation}
The same analysis can be performed for $\tilde{F}_1(z,\zb)$, for which we find the following bosonic CFT data
\begin{align}
\begin{split}
(b_1^2)_{n,s} &= 2 \lambda_{222} a_2\,  \partial_{\Dh} b_{f_{2,0}}^{(1,n)} (\Dh,s)\,, \\
( b_1^2 \gamma_1)_{n,s} &= 2 \lambda_{222} a_2\, b_{f_{2,0}}^{(1,n)} (\Dh,s)\,,
\end{split}
\end{align}
and the resummation gives a compact expression:
\begin{equation}
\tilde{F}_1 (z,\zb) = - \lambda_{222} a_2 \frac{z \zb \log z \zb}{(1-z)(1-\zb)(1-z\zb)}\,.
\label{eq:tildeF1}
\end{equation}
Finally, since the discontinuity of $F_2(z,\zb)$ vanishes \eqref{eq:discsRsym}, we simply find
\begin{equation}
\tilde{F}_2 (z,\zb) = 0 \,.
\label{eq:tildeF2}
\end{equation}

\subsection{Supersymmetrization of the correlator}
\label{subsec:supersym22}

The correlation function obtained in the previous section is \textit{not} supersymmetric, i.e. the three $R$-symmetry channels given in \eqref{eq:tildeF0}, \eqref{eq:tildeF1} and \eqref{eq:tildeF2} do not respect the Ward identities given in \eqref{eq:WI}. This happens because the inversion formula misses contributions from low-lying spins $s \le s_*$ as anticipated in section \ref{subsec:LIF}.\footnote{ Such a phenomenon has already been observed for bulk correlators.
For example, in the bootstrap of the Wilson-Fisher fixed point there is an ambiguity captured by a single $\ell=0$ block \cite{Alday:2016jfr,Alday:2017zzv}.
In supersymmetric theories, one expects the inversion formula to converge better than in non-supersymmetric ones, see \cite{Lemos:2021azv} for a recent discussion.
}
The value of $s_*$ is related to the behavior of the two-point function in the Regge limit $z/\zb \to 0$ \cite{Lemos:2017vnx}, and in principle $s_*$ can be determined by careful analysis of the corresponding Witten diagrams.
Instead, in the present work we use the heuristic that $s_*$ should take the minimal value that generates a supersymmetric correlator.
As we show below, the resulting correlators make predictions which are in perfect agreement with the expectations from the topological sector.

As we just argued, in order to obtain a supersymmetric correlator, we add defect families with operators of dimensions $\Dh = 0, 1, 2, \ldots$ and low spin $s \le s_*$.
The OPE coefficients of these operators are unknowns that we fix by imposing the Ward identities \eqref{eq:WI}.
We have found experimentally that the minimal ansatz consists on taking $s_*=0$ for $\tilde{F}_1(z,\zb)$ and $s_*=1$ for $\tilde{F}_2(z,\zb)$.
To be precise, we define the final correlators $F_j(z,\zb)$ as
\begin{align}
\begin{split}
F_0 (z,\zb) &= \tilde{F}_0 (z,\zb) \,, \\
F_1 (z,\zb) &= \tilde{F}_1 (z,\zb) + \sum_{n=0}^\infty \left( k_n \hat f_{n, 0}(z, \zb) + p_n \partial_{\Dh} \hat f_{n, 0}(z, \zb) \right)\, , \\
F_2 (z,\zb) &= \tilde{F}_2 (z,\zb) + \sum_{s=0,1} \sum_{n=0}^\infty \left( q_{n,s} \hat f_{n+s, s}(z, \zb) + r_{n,s} \partial_{\Dh} \hat f_{n+s, s}(z, \zb) \right)\,.
\end{split}
\label{eq:amb22}
\end{align}
As mentioned before, the free coefficients $k_n$, $p_n$, $q_{n,s}$ and $r_{n,s}$, can be fixed by requiring that the Ward identities are satisfied. In fact this fixes all the coefficients in terms of $q_{0,0}$ and $k_1$.
Note that $q_{0,0}$ corresponds to the ambiguity $f_{0,0}(z,\zb)=1$, i.e.\ the defect identity.
However, we know from the Witten diagrams analysis of section \ref{subsec:sugra} that the defect identity is given by the constant contribution $a_2^2$, and thus
\begin{align}
\label{eq:defect-identity-amb}
 q_{0,0} = a_2^2 \Big|_{O(\frac{\sqrt{\lambda}}{N^2})} \, .
\end{align}
On the other hand, the unknown $k_1$ can be determined by demanding a bulk expansion that is consistent with the observations made in section \ref{subsec:OPEdisc}, i.e.\ there should not appear anomalous dimensions for bulk operators. This means that the expansion of \eqref{eq:amb22} in the limit $z, \zb \to 1$ should take the form of a power series, without spurious $\log(1- \zb)$ terms. Since the defect expansion \eqref{eq:amb22} is natural around $z, \zb \sim 0$, this is only possible after fixing the free coefficients and resumming the correlator. We were able to do so, and the remaining spurious term reads:
\begin{equation}
F_1 (z,\zb) \sim \frac12 \left(\lambda_{222} a_2 - k_1 \right) \log(1- \zb) + \ldots
\end{equation}
This fixes the coefficient $k_1$ to be:
\begin{equation}
k_1 = \lambda_{222} a_2\,.
\end{equation}

\subsection{Final result and comparison to localization}
\label{subsec:final22}

We will now present the final result for the correlator $\vvev{\Op_2 \Op_2}$ using the input of localization for the two remaining free coefficients, namely $a_2^2$ and $\lambda_{222} a_2$. We can then obtain OPE coefficients of other protected operators which in turn can be checked against the localization data.

The constant contribution $a_2^2$ from the defect identity can be fixed using equation \eqref{eq:single-trace-localization}:
\begin{equation}
a_2^2 = \frac{\lambda}{N^2} \left( \frac{1}{8} - \frac{3}{8 \sqrt{\lambda}} + \ldots \right)
\end{equation}
while for the OPE coefficient $\lambda_{222} a_2$ we use the localization results given in equation \eqref{eq:single-trace-localization-1} and \eqref{eq:single-trace-localization}:
\begin{equation}
\lambda_{222} a_2 = \frac{\lambda}{N^2} \left( \frac{1}{\sqrt{\lambda}} + \ldots \right)\,.
\end{equation}
We thus obtain a correlator without any free coefficient left:
\begin{align}
F_0 (z, \zb) &= - \frac{\sqrt{\lambda}}{2N^2} \frac{z \zb}{(1-z)(1-\zb)} \left[ \frac{1+z \zb}{(1- z\zb)^2} + \frac{2 z \zb \log z\zb }{(1-z\zb)^3} \right]\,, \notag \\
F_1 (z,\zb) &= \frac{\sqrt{\lambda}}{N^2} \left[ \log (1 + \sqrt{z \zb}) + \frac{z \zb}{(1 - z \zb)^2} \right. \notag \\*
&\qquad \left. + \frac{z\zb \big( 5 z\zb -2 z^2\zb^2 +z^3\zb^3 - (z+\zb)(2-z\zb+z^2\zb^2) \big) \log z\zb}{2(1-z)(1-\zb)(1-z \zb)^3} \right]\,, \notag \\
F_2 (z,\zb) &= \frac{\sqrt{\lambda}}{8 N^2} \left[ -3 - \frac{2(z+\zb)}{\sqrt{z\zb}} + \frac{(z+\zb)(1+z\zb)-4z\zb}{(1-z\zb)^2} \right. \notag \\*
&\qquad \left.  + \frac{2 \big( (z+\zb)(1+z\zb)-4z\zb\big) \log (1+\sqrt{z\zb})}{z\zb} \right. \notag \\*
&\qquad \left. +\frac{z\zb \big( (z+\zb)(3-2z\zb+z^2 \zb^2) - 6 + 6z\zb - 4z^2 \zb^2 \big) \log z\zb}{ (1-z \zb)^3} \right] \, . 
\label{eq:completion22}
\end{align}
Comparing to \eqref{eq:protected_bulk}, this correlator predicts the OPE coefficient of the double-trace operator $\Op_{(2,2)}$
\begin{equation}
\lambda_{22(2,2)} a_{(2,2)} = \frac{\lambda}{N^2} \left( \frac18 - \frac{1}{2 \sqrt{\lambda}} + \ldots \right) \,,
\end{equation}
which matches the localization results given in equation \eqref{eq:double-trace-localization}.  We can also extract the defect CFT data for the protected operators:
\begin{align}
\begin{split}
b_{21}^2 &= \frac{\lambda}{N^2} \left( \frac{1}{\sqrt{\lambda}} + \ldots \right)\,, \\
b_{2(2)}^2 &= 1 + \frac{\lambda}{N^2} \left( - \frac{1}{2 \sqrt{\lambda}} + \ldots \right)\,.
\end{split}
\end{align}
The OPE coefficient $b_{21}^2$ can be compared to the direct computation (see equation \eqref{eq:bulk-defect-localization}), and we find a perfect match. The OPE coefficient $b_{2(2)}^2$ corresponds to the operator $\Oh_{(2)}$ introduced in equation \eqref{eq:defectdoubletrace} and is a prediction from our result.\footnote{The observation that $\Oh_{(2)}$ should appear in this type of correlator was first discussed in appendix A of \cite{Giombi:2018hsx}. In principle the operator $\Oh_2$ should also appear, but it can be seen from equation \eqref{eq:bulk-defect-localization} that it is not relevant at the present order.\label{footnote-gk}}

Moreover, using the superblocks described in \cite{Liendo:2016ymz,Barrat:2020vch}, the correlator above can also be used for extracting the supersymmetric CFT data for unprotected operators. 
The resulting CFT data has to be interpreted as a sum over degenerate operators, and one would need to solve a mixing problem similar to the case of $\Nm=4$ SYM without defects \cite{Aprile:2017bgs,Aprile:2017xsp,Aprile:2017qoy,Caron-Huot:2018kta}.
Below we provide a few examples, while we postpone the full analysis of the CFT data and the mixing problem to future work. In particular, the product of tree-level coefficients and anomalous dimensions for the unprotected operators at lowest twist $\Dh = s+2$ reads
\begin{equation}
\left. \hat{F}_{s+2,s} \gamma_{s+2,s} \right|_{O(\frac{\sqrt{\lambda}}{N^2})} = - \frac{3+2s}{2(1+s)} \frac{\sqrt{\lambda}}{N^2}\,.
\end{equation}
Note that here we use the notation of \cite{Barrat:2020vch}. It is also possible to obtain a closed form for the OPE coefficients of the semishort operators $(B,1)_{[1,s]}$:
\begin{equation}
\left. \hat{E}_s \right|_{O(\frac{\sqrt{\lambda}}{N^2})} = - \frac{1+s}{4(1+2s)} \frac{\sqrt{\lambda}}{N^2}\,.
\end{equation}
\section{General identical operators}
\label{sec:33and44}

In this section we extend the analysis of the previous section for general identical operators  $\vvev{ \Om_J \Om_J}$.
The calculation of $\vvev{ \Om_2 \Om_2}$ carries through almost unchanged, as will be described shortly.
As a concrete application we obtain closed-form expressions for the $J=3,4$ correlators.

\subsection{General discussion}

As discussed in section \ref{sec:preliminaries}, the correlator of interest has the form
\begin{align}
\label{eq:form-JJ-corr}
\begin{split}
 \Fm^{(J)}(z, \zb, \OR)
 & =  \left( \frac{\OR \sqrt{z \zb}}{(1-z)(1-\zb)}\right)^J 
   + \frac{J \lambda }{2^{J+2} N^2}
   + \sum_{j=0}^J \OR^j F^{(J)}_{J-j}(z, \zb)
   + O\!\left(\frac{1}{N^2} \right) \, ,
\end{split}
\end{align}
where we used the leading-order result for the one-point function \eqref{eq:single-trace-localization}.
Here we give a general prescription to obtain the functions $F^{(J)}_{j}(z, \zb)$ that contribute at order $\frac{\sqrt{\lambda}}{N^2}$.

The central idea is to reconstruct these functions using the Lorentzian inversion formula \eqref{eq:inversion}.
It was discussed in section \ref{subsec:OPEdisc} that only operators with single-trace dimension can contribute to the discontinuity of the correlator.
The bulk OPE of $\Om_J$ takes the form \cite{Liendo:2016ymz}
\begin{equation}
\Op_J \times \Op_J \sim 
  \mathds{1} 
  + \sum_{k=1}^{J} \Bm_{[0,2k,0]} 
  + \ldots \, ,
\end{equation}
where $\ldots$ contains unprotected multiplets that do not contribute to the discontinuity.
Furthermore, $\Bm_{[0,2J,0]}$ has double-twist dimension and does not contribute.
From this it follows that, at the order we are working, the discontinuity of the correlator reads
\begin{align}
\label{eq:disc-general-k}
\begin{split}
 \Disc \Fm^{(J)}(z, \zb, \OR) \Big|_{O(\frac{\sqrt{\lambda}}{N^2})}
 = \Disc \! \bigg[ 
   \left( \frac{\OR \sqrt{z \zb}}{(1-z)(1-\zb)} \right)^J
   \sum_{k=1}^{J-1} \lambda_{JJ2k} a_{2k} \Gm_{[0,2k,0]}(z, \zb, \OR)
   \bigg] \, .
\end{split}
\end{align}
The superconformal blocks capture the information of half-BPS operators exchanged in the bulk OPE $\Om_J \times \Om_J$.
They were obtained in \cite{Barrat:2020vch}, and we reproduce them here for simplicity:
\begin{align}
\label{eq:protected-superblock}
\begin{split}
 \Gm_{[0,K,0]}(z, \zb, \OR)
 & = h_{K}(\sigma ) f_{K,0}(z,\zb) 
   + \frac{(K+2)^2 K}{128 (K+1)^2 (K+3)} h_{K-2}(\sigma ) f_{K+2,2}(z,\zb) \\
 & \quad
   + \frac{(K-2) (K+2) K^2}{16384 (K-1)^2 (K+1) (K+3)}
     h_{K-4}(\sigma ) f_{K+4,0}(z, \zb) \, .
\end{split}
\end{align}
Using \eqref{eq:disc-general-k} as input to the inversion formula, it is possible to generate a series representation of the correlators $\tilde F_{j}^{(J)}(z, \zb)$.
In concrete examples, these series expansions do not satisfy the Ward identities, just as we saw for the $\vvev{\Om_2 \Om_2} $ case.
This is a result of the inversion formula not converging for low values of the transverse spin $s$.
Empirically we have found that the minimal set of additions is
\begin{align}
\label{eq:general-J-amb}
\begin{split}
 F_{j}^{(J)}(z, \zb)
 & = \tilde F_{j}^{(J)}(z, \zb) \qquad \text{for} \qquad j = 0, \ldots, J-2 \, , \\
 F_{J-1}^{(J)}(z, \zb) 
 & = \tilde F_{J-1}^{(J)}(z, \zb)  
   + \sum_{n=0}^\infty \left( 
      k_n \hat f_{n, 0}(z, \zb) 
    + p_n \partial_{\Dh} \hat f_{n, 0}(z, \zb) \right) \, , \\
 F_{J}^{(J)} (z, \zb) 
 & = \tilde F_{J}^{(J)}(z, \zb)
   + \sum_{s=0,1} \sum_{n=0}^\infty \left( 
      q_{n,s} \hat f_{n+s, s}(z, \zb) 
    + r_{n,s} \partial_{\Dh} \hat f_{n+s, s}(z, \zb) \right) \, .
\end{split}
\end{align}
Namely, we must add $s = 0$ ambiguities to $F_{J-1}^{(J)}(z, \zb)$, while we must add $s = 0,1$ ambiguities to $F_{J-1}^{(J)}(z, \zb)$.
Using the Ward identities \eqref{eq:WI}, it is possible to fix all the free coefficients $k_n$, $p_n$, $q_{n,s}$, $r_{n,s}$ except for $q_{0,0}$ and $k_1$.
As discussed around equation \eqref{eq:defect-identity-amb}, $q_{0,0}$ can be identified with the defect identity $a_J^2$ and is therefore fixed from localization.
Furthermore, if one keeps the bulk OPE coefficients $\lambda_{JJK} a_K$ arbitrary, the Ward identities also fix their relative values.
For example, the Ward identities together with \eqref{eq:general-J-amb} imply
\begin{align}
\label{eq:rel-st-ope-coeffs}
 \lambda_{JJK} a_K  \Big|_{O(\frac{\sqrt{\lambda}}{N^2})} = \frac{K}{2^{K/2}} \lambda_{JJ2} a_2  \Big|_{O(\frac{\sqrt{\lambda}}{N^2})}
 \quad \text{for} \quad  2 \le K \le J-2 \, .
\end{align}
These results are in perfect agreement with the known values \eqref{eq:single-trace-localization} at leading order at large $\lambda$.
This provides evidence that \eqref{eq:general-J-amb} is the correct prescription for the low-spin additions.

At this point, it is possible to resum the series expansion representations for $F_{j}^{(J)}$.
These resummed correlators can be expanded in a series around $z,\zb = 1$, which corresponds to the bulk conformal block decomposition.
The expansion contains spurious $\log(1-\zb)$ terms, which would imply that bulk operators get anomalous dimensions.
These anomalous dimensions should not be present at the order we are working, so canceling the spurious logarithms fixes the remaining free parameter $k_1$.
Finally, one uses $a_J^2$ and $\lambda_{JJ2} a_2$ coming from localization to write down the final correlator.
This correlation function produces all other OPE coefficients in the protected sector, and we find perfect agreement with the literature.

\subsection{Example 1: \texorpdfstring{$\vvev{\Op_3 \Op_3}$}{<< O3 O3 >>}}
\label{subsubsec:33}

Let us turn our attention to the case of  $\vvev{\Op_3 \Op_3}$.
The purpose of this example is to illustrate the general formalism that we just discussed.
Furthermore, we present concrete intermediate results to help the readers interested in reproducing our results.

\subsubsection{Inversion of single traces}

The first step is to obtain the discontinuity of the correlator, by combining \eqref{eq:disc-general-k} and \eqref{eq:protected-superblock} with \eqref{eq:expand_f20_f42_etc}.
For concreteness we look at the coefficient of $\OR^3$, or equivalently we focus on the $F^{(3)}_0(z, \zb)$ correlator:
\begin{align}
\begin{split}
 \Disc F^{(3)}_0(z, \zb)
 & = 
   \lambda_{332} a_2 
   \frac{z^{3/2} \left(z^2-2 z \log z-1\right)}{2(1-z)^5} 
   \Disc \frac{1}{\yb^2}
   \\
 & - \lambda_{334} a_4
   \frac{z^{3/2} \big(z^3+9 z^2- 9 z -1 -6 z(z+1) \log z \big)}{4(1-z)^6}
   \Disc \frac{1}{\yb} \, .
\end{split}
\end{align}
Similar expressions can be easily obtained for the other $F^{(3)}_j(z,\zb)$.
The next step is to insert the discontinuity in the inversion formula \eqref{eq:inversion}, and carry out the inversion order by order as $z \to 0$.
Each power $z^p$ induces a defect family with blocks $\hat f_{2p+s+2n,s}(z, \zb)$ and OPE coefficients given by the poles of $b(\Dh,s)$.
Similarly, each power $z^p \log z$ induces defect anomalous dimensions $\partial_{\Dh} \hat f_{2p+s+2n,s}(z, \zb)$ given by the double poles of $b(\Dh,s)$.
All the integrals that are needed are of the form \eqref{eq:basic_inversion} with $p=1,2$.
The resulting defect expansion for the lowest-lying operators is
\begin{align}
\label{eq:series-rep-33}
 \tilde F^{(3)}_0(z, \zb)
 & = \frac12 (3 \lambda_{332} a_2-\lambda_{334} a_4) \hat f_{3,0}(z, \zb)
   + \frac12 (5 \lambda_{332} a_2-\lambda_{334} a_4) \hat f_{4,1}(z, \zb) \notag \\*
 & \quad
   + \frac{2}{21} (121 \lambda_{332} a_2-73 \lambda_{334} a_4) 
     \hat f_{5,0}(z, \zb) 
   + 2 (5 \lambda_{332} a_2-3 \lambda_{334} a_4) 
     \partial_{\Dh}\hat f_{5,0}(z, \zb) \notag \\*
 & \quad
   + \frac12 (7 \lambda_{332} a_2-\lambda_{334} a_4) \hat f_{5,2}(z, \zb) 
   + \ldots 
\end{align}

Once again, the expansions for other $\tilde F^{(3)}_j(z,\zb)$ are obtained in an identical manner.
Unlike in the $\vvev{ \Om_2 \Om_2 }$ case, we were not able to express the defect CFT data in closed forms.
However, the previous calculation can be automatized with a computer, and the expansion can be generated efficiently up to high orders.

\subsubsection{Supersymmetrization of the correlator}

As for the $\vvev{ \Om_2 \Om_2 }$ case, the  $\tilde F^{(3)}_j(z,\zb)$ do not give a supersymmetric correlator because the Lorentzian inversion formula can miss low transverse-spin contributions. In order for $\tilde F^{(3)}_j(z,\zb)$ to satisfy the Ward identities \eqref{eq:WI}, we add the following $s =0,1$ contributions to our correlators
\begin{align}
\label{eq:amb-33}
\begin{split}
 F_{0}^{(3)}(z, \zb)
 & = \tilde F_{0}^{(3)}(z, \zb) \, , \qquad
 F_{1}^{(3)}(z, \zb) 
   = \tilde F_{1}^{(3)}(z, \zb) \, , \qquad \\
 F_{2}^{(3)}(z, \zb)
 & = \tilde F_{2}^{(3)}(z, \zb)  
   + \sum_{n=0}^\infty \left( 
      k_n \hat f_{n, 0}(z, \zb) 
    + p_n \partial_{\Dh} \hat f_{n, 0}(z, \zb) \right) , \\
 F_{3}^{(3)}(z, \zb)
 & = \tilde F_{3}^{(3)}(z, \zb)
   + \sum_{s=0,1} \sum_{n=0}^\infty \left( 
      q_{n,s} \hat f_{n+s, s}(z, \zb) 
    + r_{n,s} \partial_{\Dh} \hat f_{n+s, s}(z, \zb) \right) .
\end{split}
\end{align}
After this addition, the functions $F^{(3)}_j(z,\zb)$ should satisfy the Ward identities \eqref{eq:WI}, which are highly constraining.

In fact, the only coefficients that remain unfixed are $\lambda_{332} a_2$, $k_1$ and $q_{0,0}$. 
In particular, notice that the precise relation between single-trace OPE coefficients is fixed:
\begin{align}
 \lambda_{334} a_4  \Big|_{O(\frac{\sqrt{\lambda}}{N^2})} = \lambda_{332} a_2  \Big|_{O(\frac{\sqrt{\lambda}}{N^2})} \, .
\end{align}
Reassuringly, this relation is consistent with localization \eqref{eq:single-trace-localization}.
Again, we believe this is strong evidence for the ansatz \eqref{eq:amb-33} for low transverse-spin ambiguities to be correct.
In the case of $\vvev{ \Om_4 \Om_4 }$ we find a similar result involving $K=4,6$, and for general $J$ we expect equation \eqref{eq:rel-st-ope-coeffs} to hold.

\subsubsection{Fixing free parameters and final result}

At this point, we have a supersymmetric correlator depending on three free parameters.
Using the series representation of the correlator \eqref{eq:series-rep-33} it is possible to find a closed form expression in terms of rational functions and logarithms.
In particular, we find
\begin{align}
 F_{2}^{(3)}(z, \zb)
 = -\frac{3}{8} (3 \lambda_{332}a_2 + 32 k_1) 
   \tanh^{-1}\sqrt{z \zb} + \ldots \, ,
\end{align}
where $\ldots$ stand for terms which have an expansion around $z,\zb=1$ involving only integer powers.
On the other hand, $\tanh^{-1}\sqrt{z \zb}$ has an expansion with $\log(1-\zb)$ terms.
These logarithms correspond to anomalous dimensions in the bulk, which should be absent by our assumptions.
We thus conclude
\begin{align}
 k_1 = -\frac{3}{32} \lambda_{332} a_2 \, .
\end{align}
In order to fix the last two coefficients we use input from the localization results of the topological sector.
Indeed, from \eqref{eq:single-trace-localization} we have that
\begin{align}
 \lambda_{332} a_2 = \frac{\lambda}{N^2} \left( \frac{3}{2 \sqrt{\lambda}} + \ldots \right) \, , \quad
 a_3^2 
 = \frac{\lambda}{N^2} \left( \frac{3}{32}
 - \frac{3}{4 \sqrt{\lambda}}
 + \ldots \right) \, .
\end{align}
As discussed around equation \eqref{eq:defect-identity-amb}, the coefficient $q_{0,0}$ has a natural interpretation as the defect identity contribution given by $a_3^2$. We thus conclude $q_{0,0} = - \frac{3 \sqrt{\lambda }}{4 N^2}$.

The final result with no free parameters takes a reasonably simple form:
\begin{align}
\label{eq:full-corr-33}
 F_0^{(3)}(z,\zb)
 & = -\frac{6 \sqrt{\lambda}}{8N^2} 
     \frac{(z \zb)^{3/2}}{\big[(1-z) (1-\zb)\big]^{2}} \left[ 
   \frac{1 + z \zb}{(1-z \zb)^2} 
 + \frac{2 z \zb \log z \zb}{(1 - z \zb)^3}
 \right] \,, \nonumber \\
F_1^{(3)}(z, \zb)
 & = \frac{3\sqrt{\lambda}}{4N^2}
      \frac{(z \zb)^{3/2}}{(1-z) (1-\zb) (1-z \zb)^4} \bigg(
    z^2 \zb^2-38 z \zb+1 \nonumber \\*
 & \quad -\frac{2 \left((z+\zb) (z \zb+1) \left(z^2 \zb^2-11 z \zb+1\right)+z \zb (z \zb+5) (5 z \zb+1)\right) \log z \zb}{(1-z) (1-\zb) (1-z \zb)}\bigg) \, , \nonumber \\
 F_2^{(3)} (z,\zb) 
 & =-\frac{18 \sqrt{\lambda}}{N^2} 
     \frac{(z\zb)^{1/2}}{(1-z \zb)^2} \bigg(
     \frac{z \zb}{(1-z) (1-\zb)}
    -\frac{3 (1 + z \zb) \left(z^2 \zb^2+6 z \zb+1\right)}{32 (1 - z \zb)^2} \nonumber \\*
    & \quad
    -\frac{2 z \zb \left(7+10 z \zb+7 z^2 \zb^2\right) \log z \zb}{32 (1-z \zb)^3}
    +\frac{z \zb (1 + z \zb) \log z \zb}{2 (1-z) (1-\zb) (1-z \zb)}
 \bigg)\, , \nonumber \\
 F_3^{(3)} (z,\zb) 
 & = -\frac{3\sqrt{\lambda}}{4N^2} \bigg(
    1 
    - \frac{3 (z \zb)^{1/2} \left((z+\zb) \left(z^2 \zb^2+10 z \zb+1\right)-3 (z \zb+1)^3\right)}{4 (1-z \zb)^4} \nonumber \\*
& \quad + \frac{3 (z \zb)^{3/2} \left(5 z^2 \zb^2+2 z \zb +5 -3 (z+\zb) (z \zb+1)\right) \log z \zb}{2 (1 - z \zb)^5} \bigg) \, .
\end{align}
In principle, this correlation function contains information of infinitely many unprotected operators, but as for the $\vvev{\Op_2 \Op_2}$ case we leave a detailed analysis of the OPE for future work. 
Instead, we focus on the CFT data captured by the topological sector.
Comparing the full correlator \eqref{eq:form-JJ-corr} to equation \eqref{eq:protected_bulk} gives the bulk data
\begin{align}
\begin{split}
 \lambda_{334} a_4 
 & = \frac{\lambda}{N^2} \left( \frac{3}{2 \sqrt{\lambda}} + \ldots \right)\, , \quad 
 \lambda_{33(3,3)} a_{(3,3)} 
 = \frac{\lambda}{N^2} \left( \frac{3}{32}-\frac{9}{8 \sqrt{\lambda}} + \ldots \right) \, , \\
\end{split}
\end{align}
while comparing to \eqref{eq:protected_def} gives the defect data
\begin{align}
 b_{31}^2 = \frac{\lambda}{N^2} \left( \frac{27}{16 \sqrt{\lambda}} + \ldots \right) \, , \quad
 b_{32}^2 = O\left(\frac{\lambda^0}{N^2}\right) \, , \quad
 b_{3(3)}^2 = 1 + \frac{\lambda}{N^2} \left( - \frac{3}{4 \sqrt{\lambda}} + \ldots \right) \, .
\end{align}
These results are in perfect agreement with the literature, see \eqref{eq:single-trace-localization}, \eqref{eq:double-trace-localization} and \eqref{eq:bulk-defect-localization}.
Remember that $b_{3(3)}$ cannot be compared to \eqref{eq:bulk-defect-localization} because the results of \cite{Giombi:2018hsx} apply to the defect operator $\hat \Om_3$, while our result applies to the defect operator $\hat \Om_{(3)}$, see the discussion in footnote \ref{footnote-gk}.

\subsection{Example 2: \texorpdfstring{$\vvev{\Op_4 \Op_4}$}{<< O4 O4 >>}}
\label{subsubsec:44}

Let us finally consider the  $\vvev{\Op_4 \Op_4}$ two-point function.
Since the calculation is essentially identical to the previous one, we skip most of the details.
However, it is important to note that the following input from localization is necessary:
\begin{align}
 \lambda_{442} a_2 = \frac{\lambda}{N^2} \left(  \frac{2}{\sqrt{\lambda}} + \ldots \right) \, , \quad
 a_4^2 
 = \frac{\lambda}{N^2} \left( \frac{1}{16}
 - \frac{15}{16\sqrt{\lambda}}
 + \ldots \right) \, .
\end{align}
Following all the steps, which can be automatized with the help of a computer, we obtain the correlation function
\begin{align}
\label{eq:full-corr-44}
 F_0^{(4)}(z,\zb)
 & = - \frac{\sqrt{\lambda}}{N^2} 
       \frac{(z \zb)^{2}}{\big[(1-z) (1-\zb)\big]^{3}} \left[ 
   \frac{1 + z \zb}{(1-z \zb)^2} 
 + \frac{2 z \zb \log z \zb}{(1 - z \zb)^3}
 \right] \, , \nonumber \\*
 F_1^{(4)}(z, \zb)
 & = \frac{\sqrt{\lambda}}{N^2}
      \frac{(z \zb)^2}{(1-z)^2 (1-\zb)^2 (1-z \zb)^4} \bigg(
    z^2 \zb^2-38 z \zb+1 \nonumber \\*
 & \quad -\frac{2 \left((z+\zb) (z \zb+1) \left(z^2 \zb^2-11 z \zb+1\right)+z \zb (z \zb+5) (5 z \zb+1)\right) \log z \zb}{(1-z) (1-\zb) (1-z \zb)}\bigg) \, , \nonumber \\
 F_2^{(4)}(z, \zb)
 & = - \frac{3 \sqrt{\lambda}}{2N^2}
     \frac{(z \zb)^2}{(1-z)^2 (1-\zb)^2 (1-z \zb)^6} \bigg( \nonumber \\*
 & \quad + 2 (z+\zb) (z \zb+1) \left(8 z^2 \zb^2-91 z \zb+8\right)
         + 2 z \zb \left(43 z^2 \zb^2+214 z \zb+43\right) \nonumber \\*
 & \quad +\frac{(z \zb+1) (z^4 \zb^4+14 z^3 \zb^3+270 z^2 \zb^2+14 z \zb+1) \log z \zb}{1-z \zb} \nonumber \\*
 & \quad +\frac{(z+\zb) (7 z^4 \zb^4-46 z^3 \zb^3-222 z^2 \zb^2-46 z \zb+7) \log z \zb}{1-z \zb}\bigg) \, , \nonumber \\
 F_3^{(4)}(z, \zb)
 & = -\frac{\sqrt{\lambda}}{2N^2} \bigg( 
     \frac{270 (z \zb)^2 (z \zb+1)}{(1-z) (1-\zb) (1-z \zb)^4}
   - \frac{z \zb \left(2 z^4 \zb^4+229 z^3 \zb^3+438 z^2 \zb^2+229 z \zb+2\right)}{(1-z \zb)^6} \nonumber \\*
 & \quad
   + \frac{90 (z \zb)^2 \left(z^2 \zb^2+4 z \zb+1\right) \log z \zb}{(1-z) (1-\zb) (1-z \zb)^5}
   - 4 \log\big(1 + \sqrt{z \zb}\big) \nonumber \\*
 & \quad 
   - \frac{(z \zb)^2 \left(2 z^5 \zb^5-14 z^4 \zb^4+117 z^3 \zb^3+325 z^2 \zb^2+395 z \zb+75\right) \log z \zb}{(1 - z \zb)^7} \bigg) \, , \nonumber \\
 F_4^{(4)}(z, \zb) 
 & = \frac{\sqrt{\lambda}}{N^2} 
     \bigg( 
    -\frac{z+\zb}{2 \sqrt{z \zb}}+\frac{(z+\zb) (z \zb+1)-4 z \zb}{2 z \zb}
     \log \big( 1 +\sqrt{z \zb} \big) \nonumber \\*
 & \quad 
  + \frac{2 (z+\zb) (z \zb+1) \left(2 z^4 \zb^4-11 z^3 \zb^3+468 z^2 \zb^2-11 z \zb+2\right)}{16 (1-z \zb)^6} \nonumber \\*
 & \quad 
   -\frac{15 z^6 \zb^6-74 z^5 \zb^5+1397 z^4 \zb^4+924 z^3 \zb^3+1397 z^2 \zb^2-74 z \zb+15}{16 (1-z \zb)^6} \\*
 & \quad 
   +\frac{(z \zb)^2 (z+\zb) \left(z^5 \zb^5-6 z^4 \zb^4+14 z^3 \zb^3+56 z^2 \zb^2+315 z \zb+70\right) \log z \zb}{4 (1-z \zb)^7} \nonumber \\*
 & \quad 
   -\frac{(z \zb)^2 \left(4 z^5 \zb^5-28 z^4 \zb^4+189 z^3 \zb^3+245 z^2 \zb^2+385 z \zb+105\right) \log z \zb}{4 (1-z \zb)^7} \bigg) \, . \nonumber
\end{align}
Once again, let us compare the predictions of this correlator with the topological sector.
We find the bulk data
\begin{align}
\begin{split}
 & \lambda_{444} a_4 
 = \frac{\lambda}{N^2} \left(  \frac{2}{\sqrt{\lambda}} + \ldots \right) \, , \qquad 
 \lambda_{446} a_6 
 = \frac{\lambda}{N^2} \left(  \frac{3}{2 \sqrt{\lambda}} + \ldots \right) \, , \\
 & \lambda_{44(4,4)} a_{(4,4)} 
 = \frac{\lambda}{N^2} \left(  \frac{1}{16}-\frac{3}{2 \sqrt{\lambda}} + \ldots \right)\, ,
\end{split}
\end{align}
while the defect data is given by
\begin{align}
\begin{split}
 & b_{41}^2 = \frac{\lambda}{N^2} \left(  \frac{2}{\sqrt{\lambda}} + \ldots \right)\, , \quad
   b_{42}^2 = O\left(\frac{\lambda^0}{N^2}\right) \, , \\
 & b_{43}^2 = O\left(\frac{\lambda^0}{N^2}\right) \, , \qquad \; \qquad
   b_{4(4)}^2 = 1 + \frac{\lambda}{N^2} \left(  - \frac{1}{\sqrt{\lambda}} + \ldots \right) \, .
\end{split}
\end{align}
These results are in perfect agreement with the literature, except for $b_{4(4)}$, which is a new prediction from our calculation.

\subsection{A conjecture}

In the three examples considered in the present paper, the function $F_0^{(J)}(z,\zb)$ is very simple:
\begin{align}
\label{eq:conjecture-general-J}
 F_0^{(J)}(z,\zb)
 = -\frac{J \sqrt{\lambda}}{4 N^2}
   \frac{(z \zb)^{J/2}}{\big[(1-z) (1-\zb)\big]^{J-1}} \left[ 
   \frac{1 + z \zb}{(1-z \zb)^2} 
 + \frac{2 z \zb \log z \zb}{(1 - z \zb)^3}
 \right] \, .
\end{align}
It is tempting to conjecture that this relation also holds for $J>4$. 
From our bootstrap calculation, \eqref{eq:conjecture-general-J} is a result of a precise combination of the spacetime bulk blocks \eqref{eq:expand_f20_f42_etc}, the superblocks \eqref{eq:protected-superblock} and the OPE coefficients in the topological sector \eqref{eq:single-trace-localization}.
Perhaps from the point of view of the explicit holographic calculation the origin of \eqref{eq:conjecture-general-J} will be more transparent.

Let us also note the similarity between $F_{1}^{(3)}(z,\zb)$ and $F_{1}^{(4)}(z,\zb)$
\begin{align}
\begin{split}
 F_1^{(J)} (z, & \zb)
 = \frac{J \sqrt{\lambda}}{4 N^2}
      \frac{(z \zb)^{J/2}}{(1-z)^{J-2} (1-\zb)^{J-2} (1-z \zb)^4} \bigg(
    z^2 \zb^2-38 z \zb+1 \\
 & -\frac{2 \left((z+\zb) (z \zb+1) \left(z^2 \zb^2-11 z \zb+1\right)+z \zb (z \zb+5) (5 z \zb+1)\right) \log z \zb}{(1-z) (1-\zb) (1-z \zb)}\bigg) \, .
\end{split}
\end{align}
We do not have enough data points to propose a full analytic formula for any $J$, however our current results look promising. In the discussions below we speculate on what might be the best possible strategy for the future.
\section{Conclusions}
\label{sec:conclusions}

In this work we have studied the structure of two-point functions of single-trace half-BPS operators in the presence of a supersymmetric Wilson line in $\Nm = 4$ SYM theory. We used analytical bootstrap techniques in order to reconstruct the correlator at strong coupling. For operators of weight $J=2,3,4$ we obtained fairly simple results presented in \eqref{eq:completion22}, \eqref{eq:full-corr-33}, \eqref{eq:full-corr-44} which only involve logarithms and rational functions of the cross-ratios.

A natural continuation of our work is the analysis of two-point functions for arbitrary weight $\vvev{\Om_{J_1} \Om_{J_2}}$. One obvious approach is to keep pushing the algorithm presented in this paper to higher values of $J$. However there might be better strategies. In $\Nm=4$ SYM without defects, explicit closed form expressions for half-BPS operators take a particularly simple form in Mellin space \cite{Rastelli:2016nze,Rastelli:2017udc,Alday:2020dtb,Alday:2020lbp}.
It would be interesting to transform the explicit formulas presented in this paper to Mellin variables. The Mellin space approach for defect CFT was explored in \cite{Goncalves:2018fwx}, the hope being that this might be the natural language to write the most general correlator $\vvev{\Om_{J_1} \Om_{J_2}}$.

Another interesting line of research is to reproduce our results by an explicit holographic calculation using Witten diagrams.
Some observables in holographic defect CFTs have been studied perturbatively at strong coupling, for example bulk one-point functions \cite{Berenstein:1998ij, Giombi:2009ds}, bulk-defect correlators \cite{Giombi:2018hsx} and correlators localized on the defect \cite{Giombi:2017cqn,Giombi:2018qox,Beccaria:2019dws,Giombi:2020amn,Bianchi:2020hsz}.
The two-point function of bulk operators was studied to order $O(\frac{\lambda}{N^2})$ in \cite{Giombi:2012ep}, and the next order will involve the calculation of the Witten diagram \eqref{eq:witten-diagram-nontrivial}. The structural understanding presented in this work might give valuable input for this type of calculation.

The idea of reconstructing correlators starting from their discontinuity is powerful. In defect CFT it seems to be as powerful as in the case for homogeneous CFTs. For monodromy defects in the $\varepsilon$--expansion this method was already used to fully bootstrap two-point correlators of chiral fields \cite{Gimenez-Grau:2021wiv}. In this work we developed an algorithm that in principle can be used to bootstrap an infinite family of half-BPS correlators. We expect that the same method also works in related $\Nm = 4$ SYM setups, such as the non-supersymmetric Wilson line \cite{Beccaria:2019dws}, Wilson lines in more general representations of the gauge group \cite{Giombi:2020amn}, or even higher codimension defects. Furthermore, many half-BPS defects are known to exist in maximally supersymmetric theories in $d=3,4,6$, and all of them might be prime targets for the analytical bootstrap techniques used here.

\acknowledgments
We thank S.~Giombi and M.~Preti for useful correspondence.
We also thank V.~Forini, A.~Kaviraj, Z.~Komargodski, Y.~Linke, G.~Peveri, J.~Plefka, L.~Rastelli, J.~Rong, and V.~Schomerus for discussions and comments.
JB’s research is funded by the Deutsche Forschungsgemeinschaft (DFG, German Research Foundation) - Projektnummer 417533893/ GRK2575 “Rethinking Quantum Field Theory”.
AG and PL are supported by the DFG through the Emmy Noether research group ``The Conformal Bootstrap Program'' project number 400570283.  

\appendix

\section{Appendix}
\label{sec:appendix}

\subsection{Singular part of bulk blocks}
\label{sec:sing-blocks}

In this appendix, bulk blocks are studied in the limit $\yb = (1-\zb)/\sqrt{\zb} \to 0$.
These results provide the necessary input for the inversion formula in sections \ref{sec:22strong} and \ref{sec:33and44}.
The starting point are the explicit formulas for bulk blocks derived in \cite{Isachenkov:2018pef,Liendo:2019jpu}, which we reproduce here for convenience:
\begin{align}
\label{eq:bulkstblock}
f_{\Delta,\ell}(z,\bar{z}) 
& = \sum_{m=0}^\infty \sum_{n=0}^\infty \frac{4^{m-n}}{m! n!} \frac{\left( - \frac{\ell}{2} \right)_m^2 \left( \frac{2 - \ell - \Delta}{2} \right)_m}{\left( - \ell \right)_m \left( \frac{3 - \ell - \Delta}{2} \right)_m} \frac{\left( \frac{\Delta - 1}{2} \right)_n^2 \left( \frac{\Delta + \ell }{2} \right)_n}{\left( \Delta - 1 \right)_n \left( \frac{\Delta + \ell + 1}{2} \right)_n} \frac{\left( \frac{\Delta + \ell}{2} \right)_{n-m}}{\left( \frac{\Delta + \ell - 1}{2} \right)_{n-m}} \notag \\
& \quad \times (1-z\bar{z})^{\ell - 2m} 
\ _4F_3 \left( \begin{array}{c}
    -n, -m , \frac{1}{2} , \frac{\Delta - \ell - 2}{2} \\
    \frac{2 - \Delta - \ell - 2n}{2}, 
    \frac{\Delta + \ell - 2m}{2} , 
    \frac{\Delta - \ell - 1}{2}
    \end{array} ; 1 \right)  \\
& \quad \times \left[ (1-z)(1-\bar{z}) \right]^{\frac{\Delta - \ell}{2} + m + n} 
  \ _2F_1 \left(  \begin{array}{c}
    \frac{\Delta + \ell}{2} -m+n , \frac{\Delta + \ell}{2} -m+n  \\
    \Delta + \ell - 2(m-n) 
    \end{array} ; 1 - z\bar{z} \right) \, . \notag
\end{align}
After changing variables from $\zb \to \yb$ and expanding up to order $O(\yb^3)$, we obtain the following formulas:
\begin{align}
\label{eq:expand_f20_f42_etc}
 f_{2, 0}(z, \zb)
 & \sim -\yb \log z
 + \yb^3 \left(\frac{z+1}{8 (z-1)}-\frac{z \log z}{4 (z-1)^2}\right) \, , \notag \\
 f_{4, 0}(z, \zb)
 & \sim \yb^2 \left(-12 + \frac{6 (z+1) \log z}{z-1}\right) \, , \notag \\
 f_{4, 2}(z, \zb)
 & \sim \yb \left(\frac{90 (z+1)}{z-1}-\frac{30 \left(z^2+4 z+1\right) \log z}{(z-1)^2}\right) \notag \\
 & \quad + \yb^3 \left(-\frac{15 (z+1) \left(z^2-20 z+1\right)}{8 (z-1)^3}-\frac{15 z \left(z^2+7 z+1\right) \log z}{2 (z-1)^4}\right) \, , \\
 f_{6, 0}(z, \zb)
 & \sim \yb^3 \left(\frac{90 (z+1)}{z-1}-\frac{30 \left(z^2+4 z+1\right) \log z}{(z-1)^2}\right) \, , \notag \\
 f_{6, 2}(z, \zb)
 & \sim \yb^2 \left(-\frac{140 \left(11 z^2+38 z+11\right)}{3 (z-1)^2}
 + \frac{140 (z+1) \left(z^2+8 z+1\right) \log z}{(z-1)^3}\right) \, , \notag \\
 f_{8, 2}(z, \zb)
 & \sim \yb^3 \left(\frac{525 (z+1) \left(5 z^2+32 z+5\right)}{(z-1)^3}-\frac{630 \left(z^4+16 z^3+36 z^2+16 z+1\right) \log z}{(z-1)^4}\right) \, . \notag 
\end{align}
The symbol $\sim$ is a reminder that these expressions are valid up to corrections of order $O(\yb^4)$.
The same calculation can be carried out to higher orders in $\yb$, as will be necessary to extend the present work to correlators with $J > 4$.

\subsection{Conformal block normalization}
\label{sec:norm-conv}

In this appendix we review our normalization conventions for conformal blocks.
Furthermore, we study the contribution of half-BPS operators to the bulk and defect OPEs.
This allows to extract topological subsector data from correlation functions, and compare to the predictions from section \ref{sec:top-sector}.

The bulk-channel conformal block is given explicitly in \eqref{eq:bulkstblock}.
In the bulk OPE limit $z,\zb \to 1$ it goes like 
\begin{align}
 \lim_{z,\zb\to1} f_{\Delta,\ell}(z,\zb) = [(1-z)(1-\zb)]^{(\Delta-\ell)/2} (1-z \zb)^\ell  \, .
\end{align}
For the defect-channel conformal block we use the same normalization as \cite{Lemos:2017vnx}
\begin{align}
\label{eq:defect-block}
 \fh_{\Dh,s}(z, \zb)
 = z^{\frac{\Dh-s}{2}} \zb^{\frac{\Dh+s}{2}} 
 \, _2F_1\left(\frac{1}{2},-s;\frac{1}{2}-s;\frac{z}{\zb}\right) 
 \, _2F_1\left(\frac{1}{2},\Dh;\Dh+\frac{1}{2};z \zb\right) \, .
\end{align}
The asymptotics in the defect OPE limit $z,\zb \to 0$ can be easily extracted.
Finally, we use the following form of the $R$-symmetry blocks:
\begin{align}
\label{eq:R-sym-block}
\begin{split}
 & h_K(\OR)
 = \OR^{-K/2} 
   \, _2F_1\left(-\frac{K}{2},-\frac{K}{2};-K-1; \frac{\OR}{2} \right) \, , \\
 & \hat h_{\hat K}(\OR)
 = \OR^{\hat K} 
 \, _2F_1\left(-\hat K-1,-\hat K;-2 (\hat K+1); \frac{2}{\OR}\right) \, .
\end{split}
\end{align}
Using \eqref{eq:superconformalinvariance} and \eqref{eq:Rsymmetryinvariant}, it is possible to check that these blocks satisfy the appropriate Casimir equations.

In the discussion of the main text, it is crucial to compare to the predictions in the topological subsector of section \ref{sec:top-sector}.
We start showing how to extract the topological bulk CFT data from our correlators.
The choice of normalization in \eqref{eq:norm-top-2pt-3pt} fixes uniquely the bulk OPE
\begin{align}
\label{eq:bulk-OPE}
\begin{split}
 & \Om_{J}(x_1, u_1) \Om_{J}(x_2, u_2) \big|_{\Om_K}
 = 
      \lambda_{JJK}
      \frac{(u_1 \cdot u_2)^{J - \frac K2}}{(x_{12}^2)^{J-\frac K2} }
      \frac{(u_1 \cdot D^{(6)}_u)^{\frac K2}
            (u_2 \cdot D^{(6)}_u)^{\frac K2}
            \Om_{K}(x_2, u)}{K! (K+1)! } + \ldots
\end{split}
\end{align}
This is the contribution of a single half-BPS operator $\Om_K(x,u)$ to the bulk OPE, and we suppress terms subleading as $x_{12}^2 \to 0$.
In order to deal with the $R$-symmetry polarization, we use the $SO(r)$ Todorov operator
\begin{align}
\label{eq:todorov}
 (D^{(r)}_u)_\mu 
 = \left( \frac r2 - 1 + u \cdot \frac{\partial}{\partial u} \right)
 \frac{\partial}{\partial u^\mu}
 - \frac{1}{2} u_\mu \frac{\partial^2}{\partial u \cdot \partial u} \, .
\end{align}
One can insert the bulk OPE \eqref{eq:bulk-OPE} combined with the one-point function \eqref{eq:normalization-1pt} in the two-point function.
Keeping only the leading term as $z,\zb \to 1$ gives
\begin{align}
\label{eq:protected_bulk}
\begin{split}
 \Fm^{(J)}(z,\zb,\OR) \big|_{\Om_{K}}
 = \left( \frac{\sqrt{z \zb} \, \sigma }{(1-z)(1-\zb)}\right)^J
    \lambda_{JJK} a_K h_K(\OR) 
    \big[ (1-z) (1-\zb) \big]^{\frac K2} + \ldots
\end{split}
\end{align}
This is the leading contribution of a protected half-BPS operator to the bulk OPE, and it is equally valid for single- and multi-trace operators.
It is reassuring that the contribution is proportional to the $R$-symmetry block \eqref{eq:R-sym-block}.
This result also justifies our choice of the overall normalization for the bulk superblock \eqref{eq:protected-superblock}.

The story for the defect OPE works in an identical way.
The form of the correlators \eqref{eq:bulkdefecttwopt} fixes uniquely the bulk-defect expansion
\begin{align}
 \Om_{J}(x, u) \big|_{\hat \Om_{\hat K}}
 = b_{J\hat K} 
   \frac{(u \cdot \theta)^{J-\hat K}}{|x^\bot|^{J-\hat K}}
   \frac{(u \cdot D^{(5)}_{\hat u})^{\hat K} \Oh_{\hat K}(\tau, \hat u)}{\hat K! (3/2)_{\hat K}} + \ldots
\end{align}
As before, we focus on the contribution of a protected defect operator and keep only the leading-order term as $x^\bot \to 0$.
Since the defect operator transforms as an $SO(5)$ symmetric traceless tensor, we use the $r=5$ version of \eqref{eq:todorov}.
Inserting the defect OPE in the two-point function gives
\begin{align}
\label{eq:protected_def}
\begin{split}
 \Fm^{(J)}(z,\zb,\OR) \big|_{\hat \Om_{\hat K}}
 \sim b_{J\hat K}^2 \hat h_{\hat K}(\OR) (z \zb)^{\hat K/2} + \ldots
\end{split}
\end{align}
This is the leading contribution as $z,\zb \to 0$ of a protected defect operator to the defect OPE.

\bibliography{./auxi/Notes}
\bibliographystyle{./auxi/JHEP}

\end{document}